\documentclass[10pt,conference]{IEEEtran}
\IEEEoverridecommandlockouts
\usepackage{glossaries}
\usepackage{cite}
\usepackage{tikz}
\usetikzlibrary{arrows.meta,positioning,calc,fit,backgrounds}
\usepackage{verbatim}
\usepackage{pifont}
\usepackage{url}
\usepackage{amssymb,amsfonts}
\usepackage{amsmath}
\usepackage{algorithm}
\usepackage{algpseudocode}
\usepackage{graphicx}
\usepackage{tabularx}
\usepackage{textcomp}
\usepackage{xcolor}
\usepackage{fancyhdr}
\usepackage{mathtools}

\usepackage{subfigure}
\usepackage[top=0.75in, bottom=1.05in, left=0.625in, right=0.625in]{geometry}
\newacronym{RL}{RL}{Reinforcement Learning}
\newacronym{LLM}{LLM}{Large Language Model}
\newacronym{GenAI}{GenAI}{Generative AI}
\newacronym{DT}{DT}{Decision Transformer}
\newacronym{H-ODT}{H-ODT}{Hierarchical  Online Decision Transformer}
\newacronym{MIMO}{MIMO}{Multiple-Input and Multiple-Output}
\newacronym{ODT}{ODT}{Online Decision Transformer}
\newacronym{NLP}{NLP}{Natural Language Processing}
\newacronym{HRL}{HRL}{Hierarchical RL}
\newacronym{RAN}{RAN}{Radio Access Network}
\newacronym{LOS}{LOS}{Line-of-Sight}
\newacronym{NLOS}{NLOS}{Non-Line-of-Sight}
\newacronym{UE}{UE}{User Equipment}
\newacronym{BS}{BS}{Base Station}
\newacronym{QoS}{QoS}{Quality-of-Service}
\newacronym{IRCs}{IRCs}{Intelligent RAN Controllers}
\newacronym{DQN}{DQN}{Deep-Q-Network}
\newacronym{h-DQN}{h-DQN}{hierarchical Deep-Q-Network}
\newacronym{MDP}{MDP}{Markov Decision Process}
\newacronym{TTI}{TTI}{Transmission Time Interval}
\newacronym{KPI}{KPI}{Key Performance Indicator}
\newacronym{RAG}{RAG}{Retrieval Augmented Generation}
\newacronym{A-RAG}{A-RAG}{Agentic RAG}
\newacronym{SLA}{SLA}{Service Level Agreement} 
\newacronym{Open RAN}{Open RAN}{Open Radio Access Network}
\newacronym{OFDM}{OFDM}{Orthogonal Frequency Division Multiplexing}
\newacronym{DRL}{DRL}{Deep RL}
\newacronym{SINR}{SINR}{Signal-to-Interference-plus-Noise Ratio}
\newacronym{GPU}{GPU}{Graphics Processing Unit}
\newacronym{CPU}{CPU}{Central Processing Unit}
\newacronym{RAT}{RAT}{Radio Access Technology}
\newacronym{ML}{ML}{Machine Learning}
\newacronym{MAE}{MAE}{Mean Absolute Error}
\newacronym{RBG}{RBG}{Resource Block Group}
\newacronym{TDD}{TDD}{Time Division Duplex}
\newacronym{ULA}{ULA}{Uniform Linear Array}
\newacronym{RRS}{RRS}{Radio Resource Scheduler}
\newacronym{eMBB}{eMBB}{enhanced Mobile Broadband}
\newacronym{URLLC}{URLLC}{Ultra Reliable Low Latency}
\newacronym{BE}{BE}{Best Effort}
\newacronym{6G}{6G}{Sixth-Generation}
\newacronym{AI}{AI}{Artificial Intelligence}
\newacronym{5G}{5G}{Fifth-Generation}
\newacronym{RB}{RB}{Resource Block}
\newacronym{PDCCH}{PDCCH}{Physical Downlink Control Channel}
\newacronym{FIFO}{FIFO}{First-in-First-Out}
\newacronym{CSI}{CSI}{Channel State Information}
\newacronym{CQI}{CQI}{Channel Quality Index}

\makenoidxglossaries

\def\BibTeX{{\rm B\kern-.05em{\sc i\kern-.025em b}\kern-.08em
    T\kern-.1667em\lower.7ex\hbox{E}\kern-.125emX}}

\newboolean{showcomments}
\setboolean{showcomments}{true}
\ifthenelse{\boolean{showcomments}}
{
}

\newcommand*{\affmark}[1][*]{\textsuperscript{#1}}

\begin{document}

\title{\vspace{-4mm} Reimagining RAN Automation in 6G: \\ An Agentic AI Framework with Hierarchical Online Decision Transformer \\ \vspace{-4mm}}
\vspace{-5pt}

\author{\IEEEauthorblockN{Md~Arafat~Habib\affmark[1], Medhat Elsayed\affmark[2], Majid Bavand\affmark[2], Pedro Enrique Iturria Rivera\affmark[2], \\ Yigit Ozcan\affmark[2], and  Melike Erol-Kantarci\affmark[1], \IEEEmembership{
Fellow,~IEEE}}
\IEEEauthorblockA{\affmark[1]\textit{School of Electrical Engineering and Computer Science, University of Ottawa, Canada}}\affmark[2]\textit{Ericsson Inc., Ottawa, Canada}\\
Emails:\{mhabi050, melike.erolkantarci\}@uottawa.ca,\\ \{medhat.elsayed, majid.bavand, pedro.iturria.rivera, yigit.ozcan\}@ericsson.com \vspace{-1em}
\vspace{-5pt}
}

\maketitle

\thispagestyle{fancy}   
\fancyhead{}                
\cfoot{}
\renewcommand{\headrulewidth}{0pt} 

\begin{abstract}
In this paper, we propose an Agentic Artificial Intelligence (AI) framework for wireless networks. The framework coordinates a pool of AI agents guided by Natural Language (NL) inputs from a human operator. At its core, the super agent is powered by a Hierarchical Online Decision Transformer (H-ODT). It orchestrates three categories of agents: (i) inter-slice, intra-slice resource allocation agents, (ii) network application orchestration agents, and (iii) self-healing agents. The orchestration takes place with the help of an Agentic Retrieval-Augmented Generation (RAG) module that integrates knowledge from heterogeneous sources. In this proposed methodology, the super agent directly interfaces with operators and generates sequential policies to activate relevant agents. The proposed framework is evaluated against three state-of-the-art baselines, showing improved throughput, reduced network delay, and higher energy efficiency at both slice-level and system-wide performance metrics. Also, the proposed Agentic framework introduces a bi-level human operator intent validation methodology, both at the slice-level and Key Performance Indicator (KPI)-level using generative AI-based time series predictors. We could rule out performance-degrading operator intents with an accuracy of $88.5\%$. Lastly, while being interrupted by any performance-degrading events, the self-healing capability of Agentic AI in our framework automatically recovers $90\%$ of its previous performance, avoiding quality-of-service drifts when there is no human involvement.    
\end{abstract}

\vspace{-15pt}

\section{Introduction}
\label{s1} 

The \gls{6G} mobile networks is expected to deliver unprecedented performance, including high throughput, millisecond-scale latency, extreme reliability, and pervasive intelligence across heterogeneous infrastructures. These ambitious targets exceed the capabilities of traditional rule-based and manually operated control systems, which struggle to adapt in real-time to rapidly changing traffic demands, diverse service requirements, and dynamic network topologies~\cite{N1}. Recent advances in \glspl{LLM} have opened new opportunities for improving human-machine interaction in telecommunication systems. \glspl{LLM} enable operators to specify high-level intents directly in natural language, removing the need for predefined \gls{SLA} mappings or static policy rules. This improves flexibility, reduces configuration overhead, and allows non-expert users to interact using domain-agnostic expressions. However, relying solely on human-provided intents is not feasible, as continuous operator supervision is impractical in today's traffic-intensive network deployments.

Most recent works on intent-driven network management do not examine how the network responds to performance-degrading events after deployment. For example, the methodologies proposed in ~\cite{N2,N3,N4,N10} perform intent execution through RAN application orchestration, but the optimization process is activated only when a human operator is involved. These works do not address how the system handles \gls{QoS} drift during hours when operators are unavailable. To ensure continuous reliability under dynamic conditions, self-driven intent generation and autonomous self-healing mechanisms are therefore essential.

The existing literature on intent-driven network management typically follows a four-stage workflow consisting of intent processing, validation, execution, and assurance~\cite{N3,N4,N10,N11}. While significant progress has been made in interpreting natural language intents and translating them into network policies, a major limitation remains in the execution stage. Most studies focus on a narrow class of optimization, such as resource allocation~\cite{N2,N12,N13} or RAN application orchestration~\cite{N3}. This limited execution diversity restricts the system’s ability to support the wide range of intents operators may issue. Satisfying such intents requires coordinated activation of multiple modules, which existing approaches rarely support. Moreover, although many works include an assurance phase to verify intent fulfillment, they typically lack corrective mechanisms when performance degradation occurs.

Agentic \gls{AI} offers a promising solution to these challenges by integrating modular autonomous agents with reasoning and learning capabilities~\cite{N8}. Agentic AI enables operators to express goals in natural language while allowing agents to translate these goals into coordinated actions across heterogeneous tasks. Beyond execution, agents can reason, plan, and adapt to evolving conditions without continuous human intervention~\cite{N9}. This enables zero-touch network management, where operations such as resource allocation, traffic steering, and performance recovery are handled autonomously. Self-healing agents further enhance resilience by detecting and mitigating performance degradation proactively, without requiring any human presence. Through persistent intent awareness and coordinated multi-agent control, Agentic \gls{AI} establishes a scalable and adaptive management layer suitable for the complexity of \gls{6G} \glspl{RAN}~\cite{N7}.

In this paper, we propose an Agentic \gls{AI}-based network management framework comprising five core modules. First, a super agent interfaces directly with the human operator and accepts intents in natural language. It is built on a parameter-efficient fine-tuned \gls{LLM} trained on a custom dataset to ensure high intent-processing accuracy. The super agent generates policies to orchestrate multiple agent-based modules for both intent fulfillment and emergency mitigation using a \gls{H-ODT}~\cite{N32}. Second, an Agentic \gls{RAG} module retrieves knowledge from multiple specialized databases to enable informed and autonomous decision making~\cite{N14}. Third, a \gls{RAN} slicing module employs inter-slice and intra-slice \gls{RL} agents for resource allocation. Fourth, a \gls{RAN} application orchestration module coordinates network optimization functions such as traffic steering, cell sleeping, beamforming, power control, and handover management. Finally, a self-healing module continuously monitors network performance, detects \gls{QoS} drift, and autonomously triggers corrective actions through the super agent.

The contributions of this study can be summarized as follows: 

\begin{itemize}
    \item This work proposes an Agentic \gls{AI} solution where a super agent intelligently initiates other agent-based modules to handle a diverse range of intents. For the first time in the wireless communication literature on Agentic \gls{AI}, we propose \gls{H-ODT} to introduce the required intelligence that enables the super agent to select an optimal policy based on operator intent.   
    \item Unlike traditional \gls{RAG}, which retrieves information once for a static query, we use \gls{A-RAG} that uses a 1-bit \gls{LLM} for reasoning and planning for fast query reformulation and re-ranking. This cooperative design enables dynamic multi-step retrieval and more accurate synthesis of results in intent-driven network management.
    \item Predictive validation of operator intents is performed in both the slice and \gls{KPI}-level. Previous works perform only \gls{KPI}-centric intent evaluation. 
    \item The proposed framework also incorporates a self-healing capability made possible by its Agentic nature. Through continuous monitoring of performance metrics, detection of intent drift, and autonomous execution of corrective actions, the system ensures sustained intent satisfaction and reliable network operation over time.
    \item We surpass the existing intent-driven network management framework by proposing a system capable of autonomously observing network performance degradation and executing corrective actions when no human operator is present.
\end{itemize}

Experimental results show that the proposed framework outperforms three baselines: a non-Agentic heuristic approach that selects network modules based on marginal gain per unit cost, and two Agentic baselines that retain the same Agentic architectural design but replace the proposed \gls{H-ODT} with either a \gls{HRL} agent or an offline \gls{DT} without online adaptation. The proposed method improves average system throughput by up to $32.9\%$, reduces network delay by as much as $60.9\%$, and achieves up to a threefold improvement in energy efficiency compared to both learning-based and non-Agentic baselines. Also, the proposed framework delivers consistent slice-level \gls{QoS} improvements under increasing user density and traffic load. It simultaneously increases \gls{eMBB} throughput, reduces \gls{URLLC} latency, and significantly improves tail-user performance. In addition, the framework introduces a bi-level human operator intent validation mechanism, where natural-language intents are verified at both the slice level and the KPI level using \gls{GenAI}-based time-series predictors, enabling the rejection of performance-degrading intents with an accuracy of $88.5\%$. Finally, under performance-degrading events, the self-healing capability of the Agentic AI restores approximately $90\%$ of the system’s pre-event performance, effectively preventing \gls{QoS} drifts and maintaining \gls{SLA} compliance.

The remainder of this paper is organized as follows. In Section II, we present a comprehensive review of the literature associated with our proposed network management scheme in this paper. Section III introduces the system model and problem formulation. Section IV details the proposed methodology, and Section V presents experimental results demonstrating the effectiveness of the proposed framework. Finally, Section VI concludes the paper.      

\section{Related work}
\label{s2}

We divide this section into three parts. First, we summarize the existing literature on AI-based intent-driven network management schemes. Next, we provide an overview of the existing works on Agentic \gls{AI} in the wireless domain, and lastly, we provide the literature on the decision transformer, which is a core technique used in our Agentic \gls{AI} framework in this paper. 

\subsection{\gls{AI}-based Intent-driven Network Management Schemes}
Intent-driven networking bridges the gap between user service demands and the underlying network operations. According to Nijah et al. \cite{N15}, \gls{NLP}-driven approaches provide the greatest adaptability for expressing intents in natural language. With the advent of \glspl{LLM}, \gls{NLP} has undergone a major transformation. These models outperform traditional techniques in both versatility and accuracy. Their capability to handle zero-shot, few-shot, and fine-tuned learning makes them the state-of-the-art for intent recognition and interpretation. Most existing studies on intent-driven network management employ \glspl{LLM} to comprehend operator intents. Although some generative \gls{AI} paradigms investigate diffusion models or neuro-symbolic reasoning, \glspl{LLM} remain the dominant focus in current intent-driven network management research.

Two recent studies investigate the use of \gls{LLM}-assisted techniques for intent-based management in \gls{5G} core networks. Manias et al. \cite{N16} propose an \gls{LLM}-driven framework for intent extraction leading to zero-touch network service management. Their customized \gls{LLM} interprets and converts user intents into executable network policies, minimizing human involvement. Semantic routing has been proposed in \cite{N17} by the same authors to enhance LLM-assisted intent-based networking. Compared to the conventional \gls{LLM}-driven frameworks, the work presented in \cite{N17} works on overcoming challenges such as hallucinations, limited scalability, and reduced accuracy when processing complex network intents.

The framework presented in \cite{N18} introduces a collaborative multi-agent architecture for managing shared network resources in \gls{6G}. In this system, LLM-based agents represent distinct business entities that negotiate service-level goals, including throughput, cost efficiency, and energy optimization. Acting as a central mediator, the framework employs \glspl{LLM} in combination with optimization methods and real-time network observability to resolve conflicts and maintain balanced resource allocation. Another work \cite{N19} presents a holistic \gls{LLM}-based intent life-cycle management framework that manages every stage of intent processing, including decomposition, translation, negotiation, activation, and assurance.

Chen et al. present a vision model for \gls{6G} networks built on intent-driven autonomous intelligence to enable seamless collaboration between humans and machines \cite{N20}. The authors of \cite{N20} introduce the concept of intent-driven cooperative intelligent clusters, in which heterogeneous devices dynamically form \gls{AI}-managed groups to accomplish complex tasks.

Ouyang et al. propose an intent-driven end-to-end network orchestration framework for \gls{6G} systems that automates life-cycle management of network resources across \gls{RAN}, transport, and core domains \cite{N21}. The authors design an architecture composed of intent, orchestration, knowledge, and infrastructure layers, where natural language intents are translated into network policies using \gls{NLP} and \gls{DRL}. 

While intent processing remains fundamental, verifying intent feasibility and potential impact is equally critical. The study presented in \cite{N4} introduces a transformer-based time series predictor for intent validation prior to execution. This predictive module uses historical network data to forecast traffic trends, ensuring that intended optimizations, such as enhancing energy efficiency or increasing throughput, do not degrade service quality. Once validated, an \gls{HRL}-based framework activates suitable optimization applications, including beamforming, traffic steering, and power control. An attention-enhanced \gls{HRL} model further eliminates suboptimal actions to minimize computational cost while maximizing performance.

Another significant advancement in end-to-end \gls{AI}-enabled automation is the integration of multi-agent learning architectures \cite{N11}, in which \gls{AI}-driven agents dynamically negotiate and resolve conflicting intents. This is particularly vital in multi-tenant \gls{6G} environments, where diverse stakeholders such as network operators, service providers, and enterprises compete for limited resources.

\subsection{Agentic \gls{AI} in Mobile Communication System}

Agentic \gls{AI} can be a game-changing tool in the upcoming \gls{6G} networks. Considering the network automation capabilities of Agentic \gls{AI}, researchers recently have investigated its potential in \cite{N21,N22,N23}. Following the trend, Elkael et al. propose AgentRAN, an AI-native and Open-\gls{RAN}-aligned architecture that utilizes hierarchical intent decomposition across protocol layers, time scales, and spatial domains \cite{N26}. In \cite{N51}, a conceptual and architectural foundation for Agentic \gls{AI} in \gls{RAN} management is presented.

Compared to these recent technical works in Agentic \gls{AI}, the proposed framework achieves higher intelligence and autonomy. The \gls{H-ODT} introduces predictive and goal-aware orchestration instead of static \gls{LLM} reasoning. The Agentic \gls{RAG} enhances adaptability through multi-tier reasoning-based retrieval. The framework supports diverse intents and coordinated execution across multiple modules rather than focusing only on RAN control. It also enables self-healing and zero-touch operation.

\begin{table*}[!t]
\centering
\caption{Feature Comparison Between the Proposed Framework and Existing Works in Intent-Driven Network Management}
\renewcommand{\arraystretch}{1.2}
\setlength{\tabcolsep}{3.5pt}
\begin{tabular}{lccccc}
\hline
\textbf{Ref.} & 
\textbf{\begin{tabular}[c]{@{}c@{}}Decision intelligence among \\ multi-objective agents\\ (H-ODT)\end{tabular}} & 
\textbf{\begin{tabular}[c]{@{}c@{}}Agentic RAG\\ (multi-tier retrieval)\end{tabular}} & 
\textbf{\begin{tabular}[c]{@{}c@{}}Diversity of intents \&\\ execution modules\end{tabular}} & 
\textbf{\begin{tabular}[c]{@{}c@{}}Self-healing /\\ Zero-touch \\ autonomy\end{tabular}} & 
\textbf{\begin{tabular}[c]{@{}c@{}}Multi-predictor\\ time-series forecasting\end{tabular}} \\ \hline

\cite{N26} & No & No & Limited to infrastructure orchestration & No & No \\

\cite{N25} & No & No & Moderate (RAN functions only) & No & No \\

\cite{N3} & No & No & Moderate (RAN applications) & No & No \\

\textbf{Proposed} & 
\textbf{Yes} & 
\textbf{Yes} & 
\textbf{Yes (multi-domain agentic orchestration)} & 
\textbf{Yes} & 
\textbf{Yes} \\

\hline
\end{tabular}
\label{tab1}
\end{table*}

\subsection{Decision Transformer for Wireless Networks}

A fairly new concept, decision transformer, \cite{N27} which combines reinforcement learning, supervised learning, and sequence modeling to enable goal-directed, data-efficient, and stable learning has been used successfully to optimize wireless networks \cite{N3}, \cite{N28}, \cite{N29}, \cite{N30}, \cite{N31}. The \gls{DT} architectures used in the mentioned works have a key limitation when applied to real-world, interactive settings. Since the \gls{DT} is trained purely on offline datasets, it cannot explore beyond the trajectories it has seen and is heavily dependent on data quality and prone to distributional bias. Moreover, its deterministic policy limits adaptability and prevents efficient exploration during deployment. In this work, we adopt \gls{H-ODT} proposed for the first time to the best of our knowledge in this paper on top of the main \gls{ODT} work proposed in \cite{N32}.  The proposed \gls{H-ODT} addresses the mentioned limitations of typical \gls{DT} architectures by blending offline pretraining with online finetuning in a unified framework. It introduces stochastic policies for exploration, sequence-level entropy regularization to balance exploration and exploitation, and a trajectory-level replay buffer for continual policy updates. Furthermore, hierarchy in \gls{ODT} enables the system to focus on achieving target objectives extracted from intents provided by the operator or even auto-generated by the super agent in the hours when no human is present. 

To show the difference of the work proposed in this paper compared to the most recent literature that involves Agentic \gls{AI} and intent-driven network management, we provide Table \ref{tab1}.

\section{System Model and Problem Formulation}
\label{s3}

\subsection{Network Model}

We consider a massive millimeter-wave \gls{MIMO} cellular network with a set of cells 
$\mathcal{C}=\{1,2,\ldots,C\}$ and a target cell $c_{\text{target}}\in\mathcal{C}$. 
The target cell is equipped with $N_t$ transmit antennas operating at carrier frequency $f_c$ and supports 
$\mathcal{S}=\{1,2,\ldots,S\}$ \gls{RAN} slices. A set of \glspl{UE} 
$\mathcal{U}=\{1,2,\ldots,U\}$ is served, where each \gls{UE} $u\in\mathcal{U}$ is associated with exactly one slice 
$s\in\mathcal{S}$ and has $N_u$ receive antennas.

The system bandwidth is $B$~MHz and is divided into \glspl{RB}, which are grouped into $R$ \glspl{RBG}. 
An \gls{RBG} is the minimum radio resource allocation unit. Time is discretized into \glspl{TTI} of duration 
$t_{\mathrm{TTI}}$, and each scheduling step $n$ corresponds to one TTI, i.e., $t_n = n\,t_{\mathrm{TTI}}$. 
An inter-slice \gls{RRS} allocates RBGs among slices, while an intra-slice \gls{RRS} assigns the allocated RBGs 
to individual UEs within each slice. The system operates in \gls{TDD} mode. A wideband \gls{CQI} model is adopted, such that the spectral efficiency $SE_u(n)$ of UE $u$ is identical across all \glspl{RBG} within a step.

We consider a hexagonal layout with $C=7$ cells. Performance evaluation is conducted at the target cell while accounting for inter-cell interference from the six neighboring cells. Each cell employs a three-sector base station. Each sector uses a two-dimensional antenna array formed by stacked \glspl{ULA}, enabling azimuth and elevation beamforming through horizontal steering and electrical down-tilt control.

Under RAN slicing, the inter-slice scheduler distributes the $R$ available \glspl{RBG} among slices to satisfy slice \gls{QoS} intents. Let $\mathbf{R}_n = [R_1(n), R_2(n), \ldots, R_S(n)]$
denote the \gls{RBG} allocation vector at step $n$, where $R_s(n)$ is the number of \glspl{RBG} assigned to slice $s$. The 
allocation satisfies: $\sum_{s=1}^{S} R_s(n) = R$. Each slice $s$ contains a subset $\mathcal{U}_s$ of \glspl{UE} with similar traffic characteristics and identical \gls{QoS} 
requirements. The total number of possible \gls{RBG} allocation combinations is: 
$|\mathcal{R}_{\mathrm{comb}}|=\binom{R+S-1}{S-1}$.

\subsubsection*{Throughput and Buffer Model}

Let $R^u_s(n)$ denote the number of \glspl{RBG} allocated to \gls{UE} $u$ in slice $s$ at step $n$. The served data volume (in 
bits per step) is:
\begin{equation}
r_u(n) = R^u_s(n)\, B_{\mathrm{RBG}}\, SE_u(n)\, t_{\mathrm{TTI}},
\label{eq:served_throughput}
\end{equation}
where $B_{\mathrm{RBG}}$ is the bandwidth of one \gls{RBG}. To model packet transmission, the delivered data volume is rounded to an integer number of packets with packet size $PS$:
\begin{equation}
r_u(n) \leftarrow PS \cdot \left\lfloor \frac{r_u(n)}{PS} \right\rfloor .
\end{equation}

Let $b_u(n)$ denote the buffer occupancy (in bits) of UE $u$ at step $n$. The effective served throughput is:
\begin{equation}
r_u^{\mathrm{eff}}(n) = \min\{r_u(n),\, b_u(n)\}.
\label{eq:effective_throughput}
\end{equation}
The normalized buffer occupancy is:
\begin{equation}
b_u^{\mathrm{occ}}(n) = \frac{b_u(n)}{b_{\max}},
\label{eq:buffer_occupancy}
\end{equation}
where $b_{\max}$ is the maximum buffer capacity. Packets are dropped when the buffer overflows or when a packet 
latency exceeds the maximum allowable delay $l_{\max}$. The dropped data volume at step $n$ is denoted by $d_u(n)$.

Let $a_u(n)$ denote the traffic arrivals (in bits per step). The packet loss rate over a sliding window of length 
$m$ is:
\begin{equation}
p_u(n) =
\begin{cases}
\dfrac{\sum_{i=n-m+1}^{n} d_u(i)}
{b_u(n-m) + \sum_{i=n-m+1}^{n} a_u(i)}, & n \ge m,\\[10pt]
\dfrac{\sum_{i=1}^{n} d_u(i)}
{b_u(1) + \sum_{i=1}^{n} a_u(i)}, & n < m.
\end{cases}
\label{eq:packet_loss_rate}
\end{equation}

The long-term average served throughput is:
\begin{equation}
g_u(n) =
\begin{cases}
\dfrac{1}{m} \sum_{i=n-m+1}^{n} r_u(i), & n \ge m,\\[8pt]
\dfrac{1}{n} \sum_{i=1}^{n} r_u(i), & n < m,
\end{cases}
\label{eq:longterm_throughput}
\end{equation}
and the fifth-percentile served throughput is:
\begin{equation}
f_u(n) =
\begin{cases}
P_{5\%}\!\left(r_u(n-m+1),\ldots,r_u(n)\right), & n \ge m,\\
P_{5\%}\!\left(r_u(1),\ldots,r_u(n)\right), & n < m.
\end{cases}
\label{eq:5th_percentile}
\end{equation}

\subsection{Slice Types}

\begin{figure}[!t]
\centerline{\includegraphics[width=0.75\linewidth]{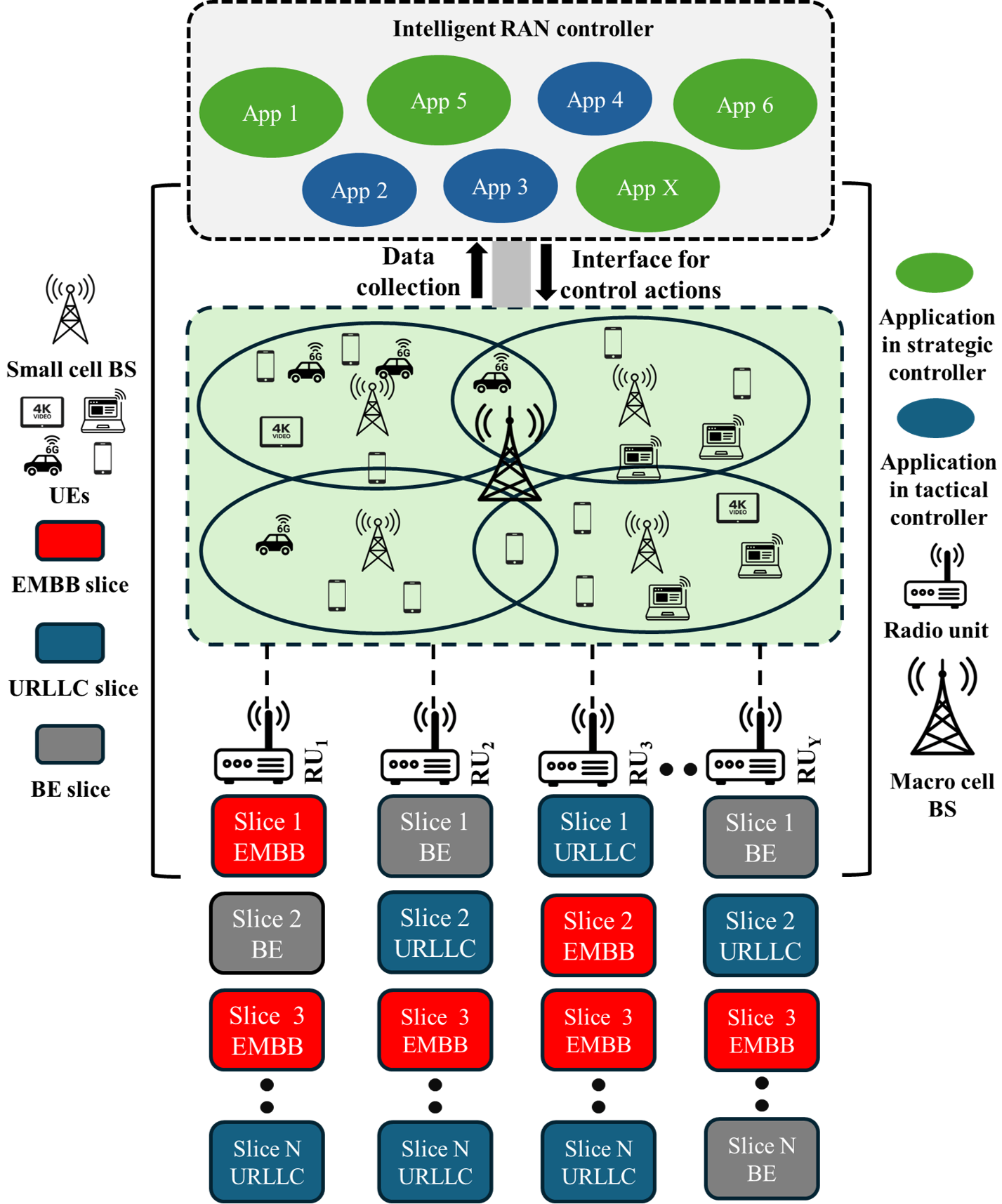}}
\caption{Network model with multiple different slices.}
\label{fig1}
\vspace{-1.2em}
\end{figure}

This work considers three slice types: \gls{eMBB}, \gls{URLLC}, and \gls{BE}. In the \gls{eMBB} slice, \glspl{UE} primarily require high throughput. Latency and packet loss constraints are comparatively relaxed. We define three \gls{QoS} requirements for \gls{eMBB}. The average served throughput $r_{\text{embb}}(n)$ must satisfy $r_{\text{embb}}(n)\ge r^{\text{req}}_{\text{embb}}$. The average latency $\ell_{\text{embb}}(n)$ must satisfy $\ell_{\text{embb}}(n)\le \ell^{\text{req}}_{\text{embb}}$. The packet loss rate $p_{\text{embb}}(n)$ must satisfy $p_{\text{embb}}(n)\le p^{\text{req}}_{\text{embb}}$. In the \gls{URLLC} slice, \glspl{UE} require ultra-low latency and high reliability. 
This is typically reflected by a very low packet loss rate. 
The throughput demand is lower than \gls{eMBB}, but the latency demand is stringent. Similar to the \gls{eMBB} case, we define \gls{QoS} constraints over throughput, latency, and packet loss using $r_{\text{urllc}}(n)$, $\ell_{\text{urllc}}(n)$, and $p_{\text{urllc}}(n)$ as the achieved metrics, and $r^{\text{req}}_{\text{urllc}}$, $\ell^{\text{req}}_{\text{urllc}}$, and $p^{\text{req}}_{\text{urllc}}$ as the corresponding target requirements. The \gls{BE} slice has the lowest priority among the three. It has no strict latency constraints. We consider two \gls{QoS} requirements for the \gls{BE} slice. The long-term served throughput $g_{\text{be}}(n)$ must satisfy $g_{\text{be}}(n)\ge g^{\text{req}}_{\text{be}}$. The fifth-percentile served throughput $f_{\text{be}}(n)$ must satisfy $f_{\text{be}}(n)\ge f^{\text{req}}_{\text{be}}$. To model intermittent traffic, \gls{BE} \glspl{UE} are switched on or off every $n_{\text{be}}$ steps. Each state occurs with probability $0.5$.

Fig. \ref{fig1} presents the network model with multiple different slices as described before. Note that the conceptual idea of the slices and the network model is inspired by the work presented in \cite{N12}. 

\subsection {Agentic System}

Agents are the core elements of Agentic \gls{AI} systems. These agents are designed to be autonomous, adaptable, and capable of reasoning and planning. They can use tools, interact with each other, and learn from their experiences over time to recover from faults. In this work, we propose a complete Agentic \gls{AI} system for \gls{RAN} management comprising four different agents.

First, a super agent performs planning, coordination, and orchestration across the other agents. It manages their goals, resolves conflicts, and aligns their actions toward a shared objective. As a higher-level controller, it maintains a system-wide view and can assign tasks or adjust strategies based on current network conditions. The proposed framework also includes a two-tier \gls{A-RAG} system to support autonomous reasoning and context retrieval for decision making. Furthermore, the proposed Agentic \gls{AI} framework includes an Agentic module with an inter-slice agent that governs how radio resources are distributed across network slices based on the system states that include slice-level performance metrics and the corresponding \gls{QoS} demands. The same module also contains intra-slice agents that assign \glspl{RBG} to individual \glspl{UE} within each slice. In addition, another Agentic module hosts multiple network applications that can directly influence network performance. When the super agent activates this module to satisfy an operator intent, \gls{HRL} agents invoke and orchestrate different \gls{RL}-based \gls{RAN} applications, such as traffic steering and beamforming. Finally, the self-healing capability is enabled by two complementary agents: a supervised learning-based \gls{KPI}-driven application selection agent and an \gls{RL}-based inter-slice self-correcting agent.

\begin{figure}[!t]
\centerline{\includegraphics[width=0.75\linewidth]{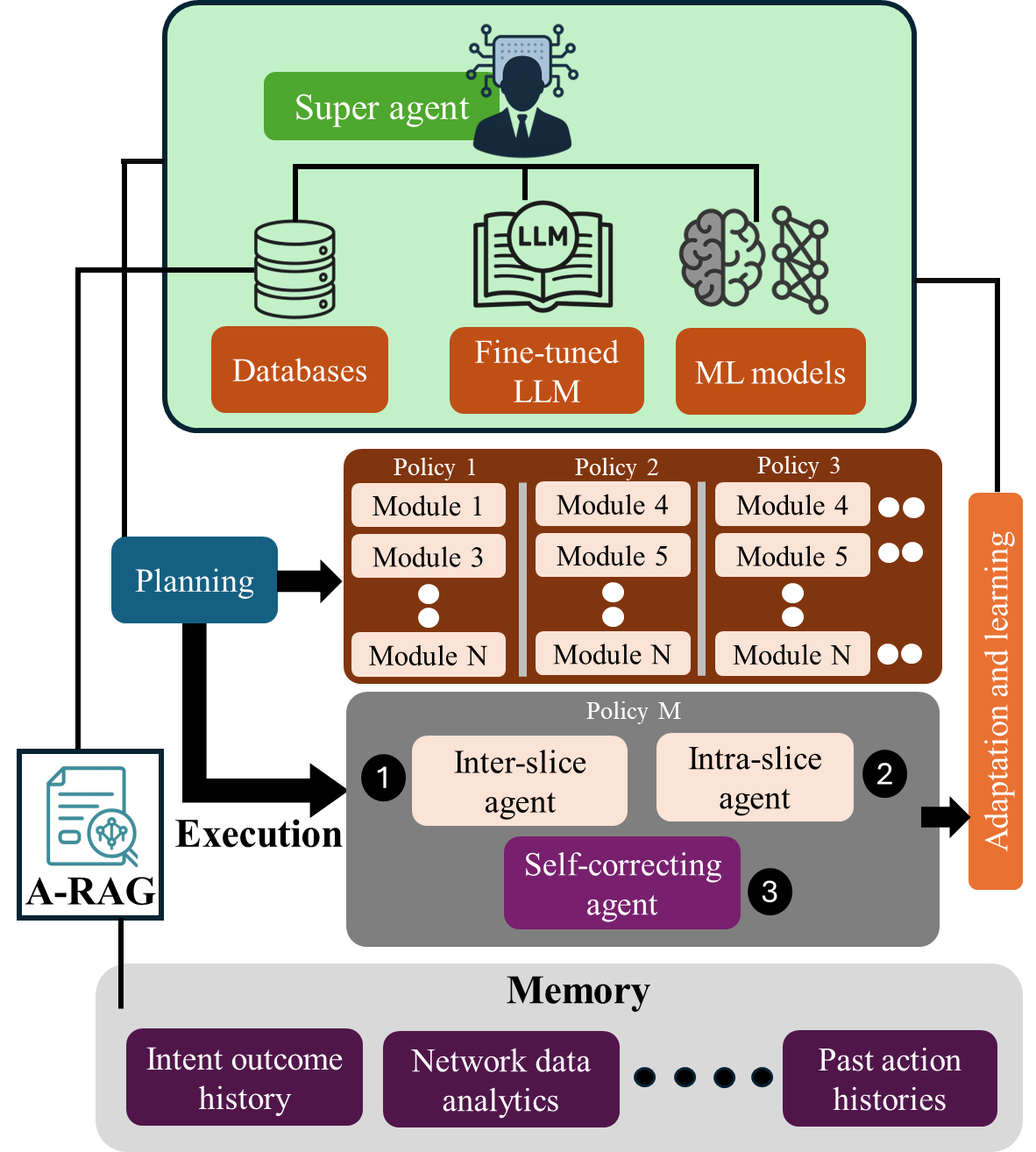}}
\caption{Agentic architecture of the system.}
\label{fig2}
\vspace{-1.2em}
\end{figure}

Fig. \ref{fig2} presents the Agentic architecture of the system. 
The super agent stays on top and is responsible for planning. The plan includes the initiation of the other agents based on the operator's intents or in a self-motivated manner. In the lower level, other agents work and interact based on the planning of the super agent.

\subsection{Hierarchical Control Architecture for \gls{RAN}}

In this work, we adopt a disaggregated Open RAN-like architecture comprising two distinct controllers: a strategic controller operating in the non-real-time domain (intervals $>1$~s), and a tactical controller functioning in the near-real-time domain (10~ms--1~s) \cite{N37}. The super agent and self-correcting agents reside within the strategic controller, responsible for long-term intent processing and corrective decision-making. The inter-slice agent is deployed at the tactical controller to handle slice-level coordination within near-real-time constraints. Meanwhile, intra-slice resource scheduling is performed at the Distributed Unit (DU) level, where low-latency, fine-grained scheduling decisions are executed per transmission interval. These agents can be deployed as a part of the O-RAN controller architecture \cite{N10} as well. From the O-RAN viewpoint, the super agent can be deployed in Service Management and Orchestration (SMO) or a non-real-time \gls{RAN} intelligent controller. The self-correcting agent can be placed in a near-real-time \gls{RAN} intelligent controller, and the intra-slice agent can be deployed in an O-RAN distributed unit.

\subsection{Problem Formulation}

Let $\mathcal{A}$ be the finite set of all agents, partitioned as
\begin{equation}
\mathcal{A}=\{\mathsf{A}_0\}\;\dot\cup\;\{\mathsf{G}\}\;\dot\cup\;\{O\}\;\dot\cup\;\{I\}\;\dot\cup\;\mathcal{J}\;\dot\cup\;\mathcal{C},
\end{equation}
where $\mathsf{A}_0$ is the super agent, $\mathsf{G}$ is the Agentic \gls{RAG}, $O$ is the \gls{RAN} application orchestration agent, $I$ is the inter-slice agent, $\mathcal{J}=\{J_s: s\in\mathcal{S}\}$ is the family of intra-slice agents, and $\mathcal{C}=\{C_s: s\in\mathcal{S}\}$ is the family of self-correcting agents. The set of decision-capable agents is defined as:
\begin{equation}
\mathcal{A}_{\mathrm{dec}} \triangleq \{O\}\cup\{I\}\cup\mathcal{J}\cup\mathcal{C}.
\end{equation}

Let $\mathcal{F}$ denote the set of abstract functionalities and let $\mathrm{cap}:\mathcal{A}\to 2^{\mathcal{F}}$ be the capability map:
\begin{equation}
\begin{aligned}
\mathrm{cap}(\mathsf{A}_0) &\supseteq \{\text{planning}, \text{coordination}, \text{orchestration}\}, \\
\mathrm{cap}(\mathsf{G}) &\supseteq \{\text{retrieval}, \text{reasoning}, \text{aggregation}\}, \\
\mathrm{cap}(I) &\supseteq \{\text{inter-slice allocation}\}, \\
\mathrm{cap}(J_s) &\supseteq \{\text{intra-slice scheduling}\}, \\
\mathrm{cap}(C_s) &\supseteq \{\text{intent-drift detection}, \text{strategy switching}\}, \\
\mathrm{cap}(O) &\supseteq \{\text{application selection and control}\}.
\end{aligned}
\end{equation}

Connectivity is modeled by a directed relation $E\subseteq \mathcal{A}\times\mathcal{A}$ with three disjoint typed sub-relations: $E_{\mathrm{ctrl}}=\{(\mathsf{A}_0,a): a\in\mathcal{A}_{\mathrm{dec}}\}$,
$E_{\mathrm{info}}=\{(\mathsf{A}_0,\mathsf{G}),(\mathsf{G},\mathsf{A}_0)\}$, and
$E_{\mathrm{analytics}}=\{(a,\mathsf{G}): a\in\mathcal{A}_{\mathrm{dec}}\}$. Thus, $\mathsf{A}_0$ invokes all decision-capable agents, $\mathsf{A}_0$ and $\mathsf{G}$ exchange information bidirectionally, and decision-capable agents report analytics only to $\mathsf{G}$. No other communication paths exist.

The global state space factorizes as $\Sigma=\Sigma_1\times\Sigma_2\times\Sigma_3$, where $\Sigma_1$ represents the radio and network state, $\Sigma_2$ encodes the orchestration state, and $\Sigma_3=\Big(\prod_{s\in\mathcal{S}}\Sigma^{(J)}_s \times \prod_{s\in\mathcal{S}}\Sigma^{(C)}_s\Big)$ captures per-slice contexts. Let $\mathbf{x}:\Sigma\rightarrow\mathbb{R}^p$ extract the $p$-dimensional \gls{KPI} vector.

Each agent $a$ applies a block-structured update $U_a:\Sigma\rightarrow\Sigma$. Informational agents satisfy $U_{\mathsf{A}_0}=U_{\mathsf{G}}=\mathrm{id}_\Sigma$. The inter-slice agent $I$ acts on $\Sigma_1$, the orchestration agent $O$ acts on $\Sigma_1$ and $\Sigma_2$, and each $J_s$ and $C_s$ act on $\Sigma_1$ and their respective slice-level context in $\Sigma_3$.

An agent impacts KPI $k$ if there exists $s\in\Sigma$ such that $\mathbf{x}(U_a(s))_k\neq\mathbf{x}(s)_k$. The resulting impact relation $\Psi\subseteq\mathcal{A}\times[p]$ satisfies:
\begin{equation}
\Psi \cap (\{\mathsf{A}_0,\mathsf{G}\}\times[p])=\varnothing, \qquad \mathcal{A}_{\mathrm{dec}}\times[p]\subseteq\Psi.
\end{equation}

Each decision-capable agent emits analytics via $G_a:\Sigma\to\mathcal{Y}$. The \gls{RAG} maintains a knowledge store $K\subseteq\mathcal{Y}$ updated as:
\begin{equation}
K_{t+1}=\mathcal{A}\!gg\!\left(K_t\cup\{G_a(s_t): a\in\mathcal{A}_{\mathrm{dec}}\ \text{executed at }t\}\right).
\end{equation}

An execution trace $(a_t,s_t)$ satisfies: (i) $a_0=\mathsf{A}_0$; (ii) $(a_t,a_{t+1})\in E$; (iii) if $a_t\in\mathcal{A}_{\mathrm{dec}}$ then $s_{t+1}=U_{a_t}(s_t)$ and $K$ is updated; (iv) if $a_t\in\{\mathsf{A}_0,\mathsf{G}\}$ then $s_{t+1}=s_t$. Any occurrence of $a_t\in\mathcal{A}_{\mathrm{dec}}$ must be bracketed by $\mathsf{A}_0$ and $\mathsf{G}$, enforcing that decision-capable agents neither call one another nor directly invoke $\mathsf{A}_0$ or $\mathsf{G}$.

The super-agent policy is
\begin{equation}
\pi_{\mathsf{A}_0}:\Sigma\times 2^{\mathcal{Y}}\to \{\mathsf{G}\}\cup\mathcal{A}_{\mathrm{dec}},
\end{equation}
selecting either information retrieval ($\mathsf{G}$) or a decision-capable agent. Zero-touch operation corresponds to $\pi_{\mathsf{A}_0}$ requiring no external inputs beyond $(s_t,K_t)$.

Let $\mathcal{I}$ be the intent set and $\Gamma:\mathcal{I}\to\mathbb{R}^p$ map intents to target KPIs. We define $\Delta_i(s)=\Gamma(i)-\mathbf{x}(s)$. Let $J_i(\mathcal{U},s)$ measure the expected reduction of $\|\Delta_i\|$ when applying $\mathcal{U}\subseteq\mathcal{A}_{\mathrm{dec}}$. The policy is consistent with intent $i$ if
\[
\pi_{\mathsf{A}_0}(s,K)\in\arg\max_{a\in\{\mathsf{G}\}\cup\mathcal{A}_{\mathrm{dec}}} \widehat{J}_i(a,s,K),
\]
where $\widehat{J}_i$ is estimated from $K$ using $\mathsf{G}$, and $a=\mathsf{G}$ is allowed for information acquisition.

The resulting sequential decision problem is:
\begin{equation}
\label{eq:cmdp}
\begin{aligned}
\max_{\pi_{\mathsf{A}_0}}\quad 
& \mathbb{E}_{\tau \sim \pi_{\mathsf{A}_0}}\!\left[
    \sum_{t=0}^{T} r_t(g,s_t,a_t)
    - \lambda \sum_{t=0}^{T} \mathcal{V}_{\mathrm{SLA}}(s_t)
  \right] \\
\text{s.t.}\quad
& a_t \in \mathcal{A}_{\mathrm{safe}}(s_t, K_t), \\
& \mathbb{E}_{\tau \sim \pi_{\mathsf{A}_0}}\!\left[\sum_{t=0}^{T} c(a_t)\right] \le C_{\max}.
\end{aligned}
\end{equation}
At each step, the super agent selects either $\mathsf{G}$ or a decision-capable agent with parameters. The reward measures progress toward the structured goal $g$, while $\mathcal{V}_{\mathrm{SLA}}$ penalizes violations. The admissible action set $\mathcal{A}_{\mathrm{safe}}$ filters unsafe or incompatible actions using $(s_t,K_t)$, and $c(a_t)$ captures orchestration overhead under budget $C_{\max}$.

The objective is to learn $\pi_{\mathsf{A}_0}$ that activates the right module at the right time while respecting the agent graph, block-structured dynamics, and SLA constraints. Classical optimization is impractical due to hybrid state-action spaces, partial observability, non-stationarity, sparse rewards, and combinatorial constraints. We therefore propose an \gls{H-ODT} framework that reformulates orchestration as a goal-conditioned sequence modeling problem, learning to predict agent activations and parameters while implicitly capturing long-horizon dependencies and constraints.

\section{Proposed Methodology}
\label{s4}

\begin{figure*}[!t]
\centerline{\includegraphics[width=0.8\linewidth]{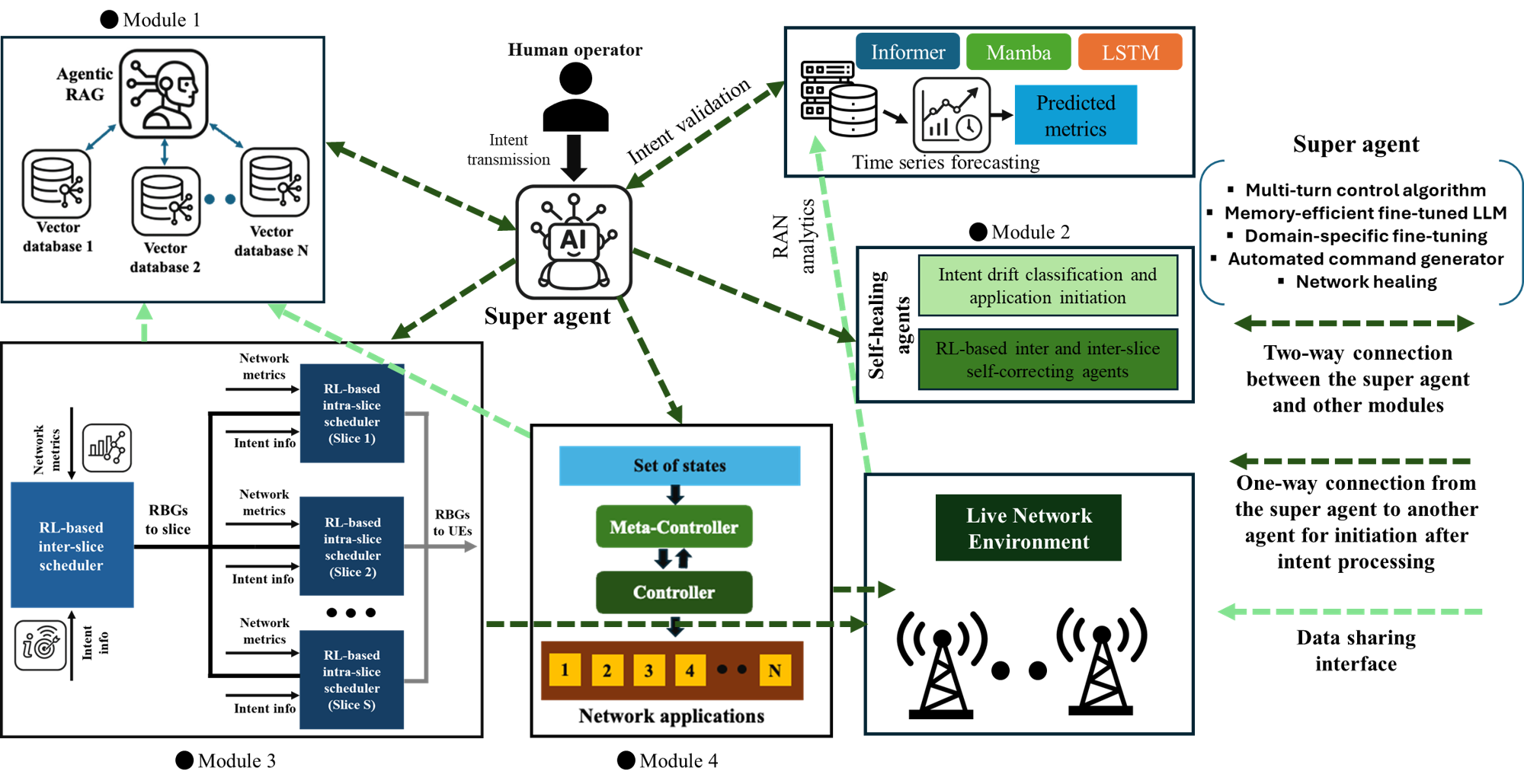}}
\caption{Proposed Agentic AI solution for autonomous network management.}
\label{fig4}
\vspace{-1.2em}
\end{figure*}

In this work, we propose an intent-driven \gls{RAN} management framework enabled by Agentic AI. As illustrated in Fig. \ref{fig4}, the proposed architecture comprises multiple specialized \gls{AI} modules with agents, each responsible for distinct decision-making functions. This section first presents the design and functionality of the super agent, which integrates a fine-tuned \gls{LLM} with an \gls{H-ODT} mechanism. Subsequently, the design and function of the remaining Agentic modules are detailed. After describing all the components, we finally present our intent-driven network management algorithm encompassing the entire process.

\subsection {Super Agent Building Blocks}

\subsubsection{\gls{LLM}-based fine-tuning}

Infused Adapter by Inhibiting and Amplifying Inner Activations (IA\textsuperscript{3}) \cite{N54} is a memory-efficient fine-tuning approach that achieves high adaptability while maintaining fast inference, which can be particularly valuable for large-scale wireless network management systems where low-latency processing is essential. Inference efficiency directly affects network responsiveness, decision latency, and energy consumption. Since IA\textsuperscript{3} trains only a small number of task-specific parameters while keeping the backbone of the pre-trained model frozen, it significantly reduces both memory and computational overhead.

In this paper, we use IA$^3$ to fine-tune an \gls{LLM} using a structured dataset comprising four distinct categories of query–response pairs: (i) intent–action reasoning, (ii) multi-database retrieval (\gls{A-RAG}), (iii) self-healing and \gls{QoS} drift, and (iv) Agentic coordination. The first category captures natural language intents that describe high-level operator objectives such as increasing throughput or improving reliability, and maps them to corresponding orchestration actions to enable the model to interpret and translate linguistic intents into domain-relevant control policies. The second category focuses on reasoning-driven retrieval tasks, where the model synthesizes responses using contextual knowledge from multiple sources such as 3GPP and Open-\gls{RAN} specifications. It ensures factual grounding and protocol awareness. The third category represents self-healing and anomaly detection scenarios that allow the model to reason about \gls{KPI} drifts and autonomously trigger corrective actions or fault conditions. Finally, the fourth category encapsulates coordination among the Agentic modules. Collectively, these dataset types ensure that the fine-tuned \gls{LLM} learns both semantic understanding and procedural reasoning required for closed-loop intent interpretation, policy orchestration, and autonomous decision-making in next-generation \gls{RAN} management. The dataset used for fine-tuning the IA$^3$-based \gls{LLM} was constructed following a structured and domain-guided approach inspired by the TSpec-LLM methodology \cite{N34}. More details on the dataset are presented in Appendix A.

\subsubsection{Agentic Retrieval Augmented Generation}

\gls{A-RAG} framework transforms the conventional retrieval-generation pipeline into an autonomous, reasoning-driven loop. In our proposed framework, a two-tier architecture enables autonomous reasoning and context retrieval for intent-driven network management. The first tier, the strategic retrieval layer, is governed by a fine-tuned super-agent that parses the operator’s natural-language intent and passes crucial parameters to the 1-bit \gls{LLM} in \gls{A-RAG} that orchestrates cross-database search. The search is performed over (i) a \gls{RAN} analytics database with real-time and historical metrics, (ii) a telecom standards database covering 3GPP and Open-RAN specifications, and (iii) an intent-history and feedback repository of prior actions and outcomes. Using hybrid multi-database retrieval and semantic re-ranking under a strict token budget, the 1-bit \gls{LLM} in \gls{A-RAG} synthesizes a compact evidence bundle and a reasoning plan for execution. The 1-bit \gls{LLM}'s \cite{N33} weights are binarized for low-latency inference while we keep embeddings, layer norms, and Key-Value (KV) cache to greater than $8$ bit for stability. Conditioned on the evidence bundle, it performs short-horizon, tool-centric reasoning to extract fine-grained domain knowledge subjected to schema and \gls{SLA} constraints enforced by the super-agent. Let $i$ denote the operator intent, and $\mathcal{D}=\{\mathcal{D}_{\mathrm{RAN}},\,\mathcal{D}_{\mathrm{STD}},\,\mathcal{D}_{\mathrm{HIST}}\}$ the \gls{RAN} analytics, standards, and intent-history corpora. The \gls{LLM} agent in \gls{A-RAG} produces a retrieval plan: $R_{\text{plan}}=\{(s_i,k_i,f_i)\}_{i=1}^m$ over sources $s_i\in\mathcal{D}$ with top-$k_i$ retrieval and filters $f_i$ and then passes it to the super agent. 

\subsection{Intent Validation}

To ensure that the execution of operator intents does not compromise network stability or performance, we employ two complementary intent validation schemes: slice-aware intent validation and \gls{KPI}-centric intent validation. The former focuses on the predicted dominance of traffic classes to prevent conflicting or counterproductive resource allocation across slices, while the later utilizes time-series forecasting of \glspl{KPI} such as throughput, packet loss, and power consumption to assess the future feasibility of an intent. Together, these two layers establish a highly robust validation framework. 

\subsubsection{Slice-Aware Intent Validation}

Let there be $C$ traffic classes indexed by $c \in \{1,\dots,C\}$. At each time step $t$, the objective of the predictor is to estimate the future composition of network traffic across these classes over the upcoming time window. Specifically, the model forecasts the next-slot class distribution as: $\hat{\boldsymbol{\mu}}_{t+1} = \big[\mu^{(1)}_{t+1}, \dots, \mu^{(C)}_{t+1}\big]$. Here, $\mu^{(c)}_{t+1}$ represents the predicted ratio of total network traffic belonging to class $c$ in the subsequent time interval. The output of the model, therefore, indicates the forecast share of each traffic class expected to dominate the network load during the next period.

For each class $c$, the model utilizes two key statistical descriptors at time $t$: $I^{(c)}_t$ (mean inter-arrival time) and $B^{(c)}_t$ (total bytes or packet volume within the observation window). These variables jointly capture the temporal and volumetric characteristics of the ongoing traffic. Shorter inter-arrival times typically indicate highly active or latency-critical traffic, whereas higher byte volumes reflect sustained high-throughput sessions. Optionally, a scalar $S_t$ can be included to represent the estimated time to the next traffic surge, obtained using a lightweight peak detection mechanism over recent load trends.  

The per-step feature vector is thus defined as: $\mathbf{x}_t = \big[I^{(1)}_t, \dots, I^{(C)}_t,\ B^{(1)}_t, \dots, B^{(C)}_t,\ S_t\big]$. The model processes a sequence of such vectors over a sliding window $\mathbf{X}_{t-W+1:t} = [\mathbf{x}_{t-W+1}, \dots, \mathbf{x}_t]$.    

In this way, we capture the temporal evolution of inter-arrival patterns and data volumes. This sequential representation enables the model to learn correlations across multiple time steps and traffic classes, allowing it to forecast not only instantaneous class proportions but also the future trajectory of traffic composition over a prediction horizon (e.g., the next several minutes).

We define a tunable dominance threshold $\tau_{\mathrm{dom}} \in (0,1)$, a minimum dominance duration $T_{\mathrm{dom}}$ to avoid oscillations, and a small hysteresis margin $h$ (e.g., two percentage points). Given an intent $I = \{c^\star, \Delta r\}$ that allocates additional resources to the slice serving class $c^\star$, the validator checks whether there exists a different class $c_{\mathrm{dom}} \neq c^\star$ such that $\mu^{(c_{\mathrm{dom}})}_{t+\tau} \ge \tau_{\mathrm{dom}} - h$ for all $\tau \in [1, T_{\mathrm{dom}}]$. If this dominance condition holds and the current QoS of $c^\star$ is already maintained within its SLA bounds (e.g., delay, loss, and throughput within target limits), the intent is invalidated to prevent starving the imminently dominant class during its surge. Otherwise, the intent is allowed, subject to the primary KPI-based validator that evaluates aggregate load, loss, and power. Typical parameter settings of $\tau_{\mathrm{dom}} \in [0.58, 0.70]$ and $T_{\mathrm{dom}}$ of a few minutes provide a practical trade-off between conservativeness and flexibility.

\begin{algorithm}[!t]
\small
\caption{Slice-Aware Intent Validator}
\label{algo1}
\begin{algorithmic}[1]
\Require Predicted mix $\hat{\boldsymbol{\mu}}_{t+1:t+T_{\mathrm{dom}}}$, intent $I=\{c^\star,\Delta r\}$
\Require Threshold $\tau_{\mathrm{dom}}$, duration $T_{\mathrm{dom}}$, hysteresis $h$
\Require QoS status of $c^\star$ (within SLA or not)
\State \textbf{Dominance check:} Find $c_{\mathrm{dom}} \neq c^\star$ such that $\mu^{(c_{\mathrm{dom}})}_{t+\tau} \ge \tau_{\mathrm{dom}} - h$ for all $\tau \in \{1,\ldots,T_{\mathrm{dom}}\}$
\If{\textbf{(i)} such $c_{\mathrm{dom}}$ exists \textbf{and} \textbf{(ii)} QoS$(c^\star)$ is within SLA}
    \State \Return \textsc{Invalidate} (to avoid starving the dominant class)
\Else
    \State \Return \textsc{Allow} (subject to the primary KPI-based validator)
\EndIf
\end{algorithmic}
\end{algorithm}

The slice-aware validator algorithm (Algorithm \ref{algo1}) evaluates whether a resource-allocation intent should be executed based on the predicted future composition of traffic classes. It checks if any traffic class other than the target class ($c^\star$) is forecast to dominate the network for a specified duration $T_{\mathrm{dom}}$, exceeding a tunable dominance threshold $\tau_{\mathrm{dom}}$. If such a dominant class exists and the \gls{QoS} of the target class is already within its \gls{SLA} limits, the intent is invalidated to prevent resource starvation of the upcoming dominant class. Otherwise, the intent is approved or passed to the primary \gls{KPI}-based validator. This simple rule ensures that resource reallocations respect short-term traffic dominance patterns and maintain overall slice stability.

\subsubsection{\gls{KPI}-Centric Predictive Intent Validation}

Before executing an operator-provided intent, it is essential to verify that the requested optimization action does not introduce adverse effects under the anticipated network operating conditions. Certain intents that are beneficial during low or moderate load regimes may lead to \gls{QoS} drift when applied under congested or unstable states. To mitigate this risk, we introduce a KPI-centric predictive intent validation mechanism that evaluates intent feasibility using short-term forecasts of key network performance indicators.

Predictive intent validation enables proactive assessment of intent feasibility by aligning optimization decisions with expected future network states. This mechanism ensures that actions such as initiating or terminating network applications remain compatible with near-future conditions, thereby preserving performance metrics including throughput, packet loss, and energy efficiency.

Let $X \in \{\text{traffic load}, \text{packet loss}, \text{power consumption}\}$ denote a predicted \gls{KPI}. For each \gls{KPI}, a historical observation sequence collected from the \gls{PDCCH} is defined as
\begin{equation}
\mathbf{X}_t = \{X(t-\Delta+1), X(t-\Delta+2), \ldots, X(t)\},
\end{equation}
where $\Delta$ denotes the observation window length. A many-to-one time-series predictor estimates the next-step KPI value as
\begin{equation}
\hat{X}(t+1) = \mathcal{F}_X(\mathbf{X}_t),
\end{equation}
where $\mathcal{F}_X(\cdot)$ represents a long-sequence forecasting model such as Autoformer, Informer, or selective state-space architectures (Mamba). The super agent selects the appropriate predictor based on sequence length, update periodicity, and computational constraints.

For each predicted KPI, adaptive upper and lower thresholds are defined. Let $\Theta = \{(U_L, L_L), (U_P, L_P), (U_E, L_E)\}$ represent threshold pairs for traffic load, packet loss, and power consumption, respectively. Using these thresholds, the predicted network state at time $t+1$ is encoded as a binary vector $\mathbf{s}(t+1) = [\sigma_L, \sigma_P, \sigma_E]$. Here,
\begin{equation}
\sigma_x =
\begin{cases}
1, & \hat{X}_x(t+1) \notin [L_x, U_x], \\
0, & \text{otherwise}.
\end{cases}
\end{equation}
This compact representation captures whether predicted KPIs violate admissible operating regions.

To evaluate intent feasibility, historical executions are analyzed to determine whether a given intent–state combination resulted in \gls{QoS} drift. A supervised feasibility table $\mathcal{T}$ maps intent types and predicted state signatures to a \gls{QoS} drift indicator:
\begin{equation}
\mathcal{T} : (i, \mathbf{s}) \rightarrow Q_{\text{drift}},
\end{equation}
where $Q_{\text{drift}} = 1$ indicates observed QoS degradation and $Q_{\text{drift}} = 0$ otherwise. An intent is considered infeasible if its predicted execution state corresponds to a positive \gls{QoS} drift entry.

The thresholds in $\Theta$ are computed dynamically by detecting significant variations in \glspl{KPI} correlated with the predicted metric. Let $M(t)$ denote the metric for which thresholds are required, and let $\text{KPI}_a(t)$ and $\text{KPI}_b(t)$ denote associated KPIs. The relative change of a \gls{KPI} is defined as:
\begin{equation}
\Delta_x(t) = \frac{\text{KPI}_x(t) - \text{KPI}_x(t-1)}{\text{KPI}_x(t-1)}.
\end{equation}
Each \gls{KPI} is associated with a significance threshold $\xi_x$, derived from historical fluctuation statistics. Two dependency types are considered: (i) increasing relationship, where \gls{KPI} growth implies degradation, and (ii) decreasing relationship, where \gls{KPI} reduction implies degradation. Algorithm 2 presents the adaptive threshold selection. 

\begin{algorithm}[!t]
\small
\caption{Adaptive Threshold Identification}
\label{alg:threshold}
\begin{algorithmic}[1]
\Require Historical tuples $(M(t), \text{KPI}_a(t), \text{KPI}_b(t))$
\Ensure Upper and lower thresholds $(U_M, L_M)$
\State Initialize $U_M \leftarrow \varnothing$, $L_M \leftarrow \varnothing$
\For{$t = 2$ to $T$}
    \State Compute $\Delta_a(t)$ and $\Delta_b(t)$
    \If{increasing relationship holds}
        \If{$\Delta_a(t) > \xi_a$ \textbf{or} $\Delta_b(t) > \xi_b$}
            \State $U_M \leftarrow M(t)$
        \EndIf
    \EndIf
    \If{decreasing relationship holds}
        \If{$\Delta_a(t) < -\xi_a$ \textbf{or} $\Delta_b(t) < -\xi_b$}
            \State $L_M \leftarrow M(t)$
        \EndIf
    \EndIf
\EndFor
\State \Return $(U_M, L_M)$
\end{algorithmic}
\end{algorithm}

Using the predicted state and the feasibility table \cite{N3}, the intent validation decision is obtained as follows using Algorithm 3. Feasibility table $\mathcal{T}$ is a supervised lookup memory that stores the empirical safety outcome (QoS drift) observed when executing an intent type $i$ under a predicted network-state signature $s$.
In other words, it implements the mapping:
\begin{equation}
\mathcal{T} : (i, s) \mapsto \bar{Q}_{\mathrm{drift}}(i,s) \in \{0,1\},
\end{equation}
where  $i$ is the intent type (Energy / Throughput / Delay / \ldots), $s = (x,y,z)$: binary network-state signature, and $\bar{Q}_{\mathrm{drift}}(i,s)$ is the binary drift label used by Algorithm~3 (0 = safe, 1 = unsafe). Algorithm~3 queries this mapping.

\begin{algorithm}[!t]
\small
\caption{KPI-Centric Intent Validation}
\label{alg:intent_validation}
\begin{algorithmic}[1]
\Require Intent $i$, predicted state $\mathbf{s}(t+1)$, feasibility table $\mathcal{T}$
\Ensure Validation decision (\textbf{Valid} or \textbf{Invalid})
\If{$(i, \mathbf{s}(t+1)) \in \mathcal{T}$}
    \If{$Q_{\text{drift}}(i, \mathbf{s}(t+1)) = 1$}
        \State \Return \textbf{Invalid}
    \Else
        \State \Return \textbf{Valid}
    \EndIf
\Else
    \State \Return \textbf{Valid}
\EndIf
\end{algorithmic}
\end{algorithm}

\subsection{Hierarchical Online Decision Transformer for Intelligent Agentic Orchestration via Super Agent}

Given an intent, the super agent generates a sequential orchestration policy that determines which agent to invoke, in what order, and with what parameters, while satisfying \gls{SLA} constraints and minimizing operational overhead. To enable long-horizon reasoning and online adaptation, we adopt a Hierarchical Online Decision Transformer (\gls{H-ODT}) framework for orchestration.

The Online Decision Transformer (\gls{ODT})~\cite{N32} reformulates reinforcement learning as a goal-conditioned sequence modeling problem and supports continual policy adaptation through online interaction and replay-based fine-tuning. Building upon this foundation, the proposed \gls{H-ODT} introduces hierarchical control and replaces manually specified returns-to-go with goal tokens derived from demonstration data and operator intents. This enables the Agentic system to directly optimize network-level objectives while maintaining online rollouts and continual finetuning under dynamic traffic and \gls{KPI} drift.

\gls{H-ODT} retains a bi-level architecture: a meta-transformer operating at a higher level and a control-transformer operating at the action level. The meta-transformer receives the recent state context and the goal token extracted from the operator intent. It produces a hierarchical conditioning token that guides the lower level. In our design, this conditioning token corresponds to an important past action $\alpha_{n-\beta}$ which previously reached the goal partially or fully (partial goal fulfillment is acceptable). This mechanism helps the low-level transformer focus on actions that are valuable for completing the task, resulting in intent-fulfilling orchestration decisions.

\begin{figure}[!t]
\centerline{\includegraphics[width=0.65\linewidth]{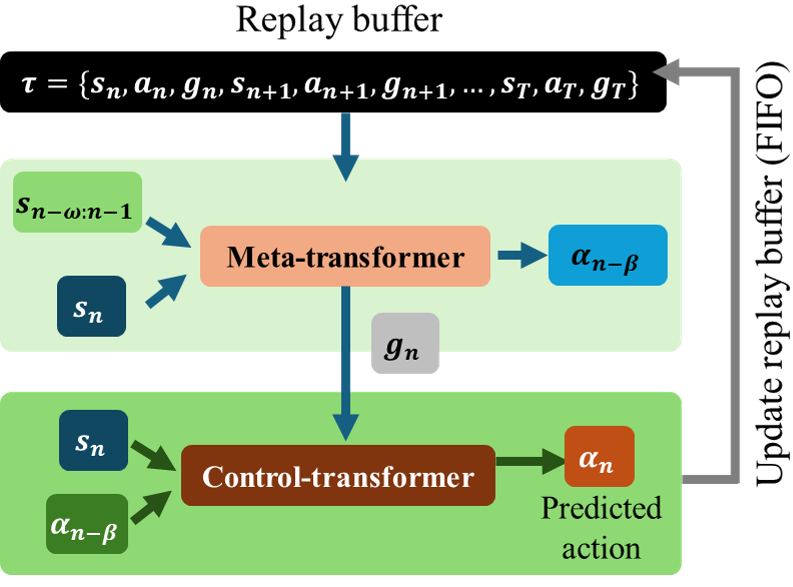}}
\caption{H-ODT architecture with rollouts.}
\label{pta}
\vspace{-1.2em}
\end{figure}

We refer to the transformer on top as the meta-transformer and the transformer on the bottom as the control-transformer. Fig.~\ref{pta} illustrates this bi-level architecture. In the figure, $s_{n-1}$ is the past state, $\alpha_{n-\beta}$ is the useful action in the past that has reached the goal partially or fully, $g_{n}$ is the goal to be achieved extracted from the operator intent, and $\alpha_n$ is the action to be taken by the control-transformer at time $n$.

At each time step $n$, the environment provides a state $s_n \in \mathcal{S}$, and the objective is to learn a goal-conditioned hierarchical policy that selects orchestration actions to reach the desired goal. In \gls{H-ODT}, the hierarchical policy is represented as:
\begin{equation}
\pi(\alpha_n \mid s_{\le n}, g_n) \;=\; \pi^{control}_{\theta}\!\big(\alpha_n \mid s_{n-K:n}, g_n, \alpha_{n-\beta}\big),
\label{eq:hodt_policy}
\end{equation}
where the meta-transformer produces $\alpha_{n-\beta}$ as:
\begin{equation}
\alpha_{n-\beta} \sim \pi_{\phi}^{meta}\big(\,\cdot \mid s_{n-\omega:n-1}, s_n, g_n\big),
\label{eq:hodt_meta}
\end{equation}
and $\omega$ is the window length of past states considered. The symbol $\sim$ emphasizes that \gls{H-ODT} supports stochastic sampling during online rollouts to enable exploration and avoid overfitting to a static offline distribution.

Each component states $s_n$, goals $g_n$, and important past actions $\alpha_{n-\beta}$ is embedded and fed into the transformer. Let $E(\cdot)$ denote the embedding function and $P(\cdot)$ the positional encoding:
\begin{itemize}
    \item State embeddings: $E(s_n) + P(n)$
    \item Goal embeddings: $E(g_n) + P(n)$
    \item Important action embeddings: $E(\alpha_{n-\beta}) + P(n-\beta)$
\end{itemize}
Thus, the transformer input token at time $n$ is:
\begin{equation}
\text{tr}_{i_n} = \left[ E(s_n) + P(n), \, E(g_n) + P(n), \, E(\alpha_{n-\beta}) + P(n-\beta) \right].
\label{eq:hodt_token}
\end{equation}
The self-attention mechanism computes attention weights across these embeddings, allowing the model to attend to goal-relevant action anchors effectively.

Unlike purely offline hierarchical decision transformers~\cite{N3}, \gls{H-ODT} continuously interacts with the environment and updates its policy online. For each validated intent, the agent executes an episode of length $T$. At every time step $n$, the meta-transformer selects a hierarchical conditioning token $\alpha_{n-\beta}$ based on the recent state context and the goal, and the control-transformer then selects an orchestration action $\alpha_n$ conditioned on the state, goal, and the selected token. The selected action is applied to the environment, producing the next system state $s_{n+1}$ and corresponding \gls{KPI} measurements. Repeating this process over $T$ time steps generates a complete rollout trajectory.

The resulting episode trajectory is represented as
\begin{equation}
\tau = \{(s_n, g_n, \alpha_{n-\beta}, \alpha_n)\}_{n=1}^{T},
\label{eq:hodt_traj}
\end{equation}
and is stored in an episode-level replay buffer $\mathcal{B}$ using a \gls{FIFO} strategy. The collected trajectories are periodically used to update the hierarchical policy by supervised learning, where the model is trained to better predict actions conditioned on observed state-goal contexts. To improve robustness under non-stationary network conditions, optional goal-consistent relabeling can be applied using observed \gls{KPI} outcomes.

Since \gls{H-ODT} conditions on goal tokens rather than manually chosen return-to-go, we maintain conditioning consistency via goal-consistent relabeling during training.  After each rollout, goal tokens can be optionally relabeled using achieved \gls{KPI} deltas (or achieved satisfaction margins) computed from the resulting trajectory, to enable stable online learning under distribution shift. Finally, the predicted action $\alpha_t$ corresponds to selecting Agentic modules (and their parameters) for initiation to fulfill the operator intent.

Every \gls{DT} fundamentally relies on reinforcement learning–generated trajectories \cite{N27}. To generate training data for \gls{H-ODT}, we employ a two-level \gls{h-DQN} framework comprising a meta-controller and a controller \cite{N53}. These controllers are specific to the \gls{h-DQN} algorithm for managing network-optimizing applications. The meta-controller and controller are hosted in the strategic and tactical controllers, respectively. The meta-controller receives the network state (e.g., traffic class) and a goal (desired change in a performance metric from an intent). The lower-level controller selects an Agentic module or a combination of them based on this input. The MDP for generating training trajectories for \gls{H-ODT} is as follows:  

\noindent
$\bullet$~\textbf{State space:} The state observed by the super agent at turn $n$ is defined as: $s_n = \{ z_n, k_n, f_n, q_n, d_n \}$. Here, $z_n$ represents the target slice. The component $k_n$ contains real-time network indicators retrieved from \gls{RAN} analytics or the \gls{A-RAG} module, including slice-level KPIs, resource utilization, and system-wide statistics. The vector $f_n$ includes short-horizon forecasts of critical metrics such as traffic load, packet loss probability, and power consumption. The variable $q_n$ captures partial intent fulfillment (e.g., percentage of KPI improvement achieved), enabling early termination or continuation decisions. Finally, $d_n$ is a discrete mode indicator that distinguishes between human-driven intent execution, autonomous operation with no active intent, and \gls{QoS} drift conditions that trigger self-healing behavior.

\noindent
$\bullet$~\textbf{Action Space:} The action selected by the super agent is hybrid in nature and defined as
\begin{equation}
a_n = \{A_{ID(n)}, \theta_n\}, \quad A_{ID(n)} \in \{A_1,\dots,A_n\}, \ \theta_n \in \Theta_{A_{ID(n)}}.
\end{equation}
Here, $A_{ID(n)}$ is a discrete identifier selecting which functional agent to invoke (e.g., \gls{A-RAG}, inter-slice scheduler, \gls{RAN} application orchestration agent, or self-healing agent), while $\theta_t$ denotes the agent-specific execution parameters. This hybrid formulation allows the super agent to reason jointly over which agent to activate and how to configure it.

\noindent
$\bullet$~\textbf{Goal for the controller:}
$g_n=\{m_n,\delta_n\}$ represents the goal for the controller to achieve. Here, $m_n$ denotes the target \gls{KPI} (e.g., throughput, delay, energy efficiency), $\delta_n$ is the desired improvement magnitude.

\noindent
$\bullet$~\textbf{Intrinsic reward design:}
The reward function is designed to promote monotonic progress toward the structured goal while enforcing continuous penalization of \gls{SLA} violations and orchestration overhead:
\begin{equation}
r_n(s_n,a_n) = 
w_m \Delta \mathrm{KPI}_m(n) 
- \lambda V_{\mathrm{SLA}}(s_{n+1})
- \eta c(a_n),
\label{eq:reward}
\end{equation}
where $\Delta \mathrm{KPI}_m(n)$ denotes the signed improvement of the target performance metric $m$ between consecutive states, defined such that positive values indicate progress toward the goal. The coefficient $w_m>0$ provides weights to the importance of the target \gls{KPI}. The term $c(a_n)$ models the operational cost of executing action $a_n = \{A_{ID(n)}, \theta_n\}$. The term $c(a_n)$ captures signaling overhead, computational load, and reconfiguration latency associated with invoking agent $A_{ID(n)}$ with parameters $\theta_n$. The nonnegative function $V_{\mathrm{SLA}}(s)$ quantifies the magnitude of \gls{SLA} violations at state $s$ by aggregating deviations of monitored KPIs from their admissible bounds. The weighting coefficients $\lambda>0$ and $\eta>0$ regulate the trade-offs among performance maximization, \gls{SLA} compliance, and orchestration efficiency. 

Goal satisfaction is defined by the set
\begin{equation}
\mathcal{C}_g = \{ s \in \mathcal{S} :
\sigma_m \big( \mathrm{KPI}_m(s) - \mathrm{Target}_m(g) \big) \ge 0 \},
\end{equation}
where $\sigma_m = +1$ for maximization-oriented KPIs (e.g., throughput, energy efficiency)
and $\sigma_m = -1$ for minimization-oriented KPIs (e.g., delay, packet loss, power).

\noindent
$\bullet$~\textbf{Extrinsic rewards:}
Summation of the intrinsic reward over $\psi$ steps is considered as the extrinsic reward. $\psi$ represents \gls{RL}-based episodic time step consisting of multiple \glspl{TTI}. Zero instantaneous reward in the intrinsic reward function corresponds to a neutral transition in which the target \gls{KPI} does not change (or its improvement is exactly offset by SLA-violation and orchestration-cost penalties). Learning remains effective because the controller optimizes the cumulative reward over $\psi$ TTIs (extrinsic reward) and \gls{H-ODT} further stabilizes training via goal-consistent relabeling under sparse feedback.

The \gls{H-ODT} is trained on a replay buffer containing Agentic orchestration trajectories: $\psi_g = [ (s_1,g_1,a_1),\dots,(s_T,g_T,a_T) ]$. These trajectories are collected from both offline simulations and online rollouts. Training is performed via supervised sequence modeling, where the transformer learns to predict the next hybrid action given the context window of past returns, states, and actions. 

\subsection{Inter-slice Agent}

The inter-slice resource allocation in the proposed Agentic \gls{AI} framework is performed by a \gls{DRL}-based inter-slice agent adopted from \cite{N12}. The agent operates at the slice level and is responsible for determining the allocation of \glspl{RBG} among active slices at each scheduling step, while respecting the system-wide resource constraint. Please refer to \cite{N12} for more details. 

\subsection{Intra-slice Agent}

For each global scheduling interval $n$, the intra-slice agent of slice $s$ receives $R_s(n)$ \glspl{RBG} from the inter-slice allocator \cite{N52}. It then assigns these resources sequentially over $T_s = R_s(n)$ micro-decisions, indexed by $\kappa \in \{1,\ldots,T_s\}$, i.e., one \gls{RBG} per micro-step. At micro-step $\kappa$, the agent observes a micro-state $s^{(s)}_{n,\kappa}$. 
This micro-state aggregates normalized features such as queue occupancies $\mathbf{q}$ and channel qualities $\mathbf{c}$. For \gls{URLLC}, it is augmented with head-of-line delays $\boldsymbol{\ell}$. For \gls{eMBB}/\gls{BE}, it includes service-share statistics $\mathbf{u}$. 
The slice budget $R_s(n)$ is also included in the state.

The action at micro-step $\kappa$ is $a^{(s)}_{n,\kappa}\in\{1,\ldots,N_s\}$, which selects the \gls{UE} to be served. Executing this action updates the slice queues, delay counters, and the cumulative \gls{RBG} allocation.

We normalize key variables to keep the state and reward bounded. 
Specifically, delay, served bits, and per-\gls{UE} share are defined as
$\tilde{\ell}_j(n,\kappa)=\ell_j(n,\kappa)/\ell_{\max}$,
$\hat{b}_j(n,\kappa)=b_j(n,\kappa)/b_{\max}$, and
$u_j(n,\kappa)=\mathrm{RBG}_j(n,\kappa)/R_s(n)$, respectively.
Let $\bar{u}(n,\kappa)$ denote the mean share across \glspl{UE} in slice $s$.

The bounded reward $r^{(s)}(n,\kappa)\in[-1,1]$ is aligned with the slice objective. 
For \gls{URLLC}, we prioritize low latency using $1-2\tilde{\ell}_j$. 
For \gls{eMBB}, we promote instantaneous throughput using $2\hat{b}_j-1$. 
For \gls{BE}, we encourage fairness by penalizing deviations from the mean share, i.e., $1-2\lvert u_j-\bar{u}\rvert$. 
This design enables latency, throughput, and fairness-aware intra-slice allocation within the Agentic hierarchy.

\subsection{\gls{RAN} Application Orchestration}

One of the best ways to improve system performance is to use DRL-based RAN applications that perform specific network modifications. We use the methodology proposed in \cite{N3} for RAN application selection and orchestration. The application pool includes a traffic steering module that employs a \gls{DQN}-based mechanism to dynamically distribute traffic between LTE and \gls{5G} NR, with decisions guided by \gls{QoS} requirements, particularly throughput and delay \cite{N35}. To improve energy efficiency, a cell sleeping application is developed, which determines the activation status of \glspl{BS} based on real time traffic load ratios and queue lengths, ensuring that active \glspl{BS} remain efficiently utilized without becoming overloaded \cite{N36}. The power allocation application focuses on maximizing total throughput by optimizing power levels for each RBG across all the \glspl{BS} \cite{N37}. In addition, a \gls{DQN}-based beamforming and power control mechanism utilizes \gls{UE} location information to select optimal beam steering angles and adjust transmission power to achieve a balance between throughput and energy efficiency \cite{N38}. Finally, an energy-efficient handover management application, also driven by \gls{DQN}, is implemented to adapt handover policies that minimize energy consumption while maintaining seamless user connectivity \cite{N39}. For example, a broader intent such as `` Improve throughput and delay by 20\%'' can lead to the invocation of a traffic steering application.   

\subsection{Self-healing Agents}

In our proposed Agentic system, we observed two important issues. First, even after launching multiple Agentic modules, the intended objective may still not be achieved. Second, unexpected network changes during hours when there is no human administrator present, (e.g., around 3 a.m.) can be challenging. For example, there can be a sudden spike in throughput demand due to an emergency. 

Our proposed system can autonomously apply corrective actions based on the network state at that timestamp to handle any increased demand. If the problem is slice-specific, it dynamically adjusts slice priorities using critical network parameters. If the intent is broader and not tied to a specific slice, the system performs supervised \gls{RAN} application selection, guided by the historical record of past application activations.

This is where the advantage of an Agentic framework becomes clear: it can self-heal to prevent severe performance degradation by invoking self-healing agents that either (i) update slice priorities or (ii) initiate and orchestrate suitable \gls{RAN} applications. The next two subsections detail these two mechanisms, which together form the self-healing component of the proposed network management system.

\subsubsection{Inter-slice Self Healing Agent}

The inter-slice self-healing agent is formulated as an \gls{MDP} with state $s_n\in\mathcal{S}$ capturing the observed conditions that drive corrective adaptation, i.e., $s_n=[D_n,\,L_n,\,\bar{A}_n]$, where $D_n=[\delta_n^{(1)},\ldots,\delta_n^{(m)}]$ collects slice-level \gls{QoS} deviations with $\delta_n^{(i)}=\mathrm{QoS}_{\mathrm{desired}}^{(i)}-\mathrm{QoS}_{\mathrm{actual}}^{(i)}$, $L_n$ denotes the instantaneous system load, and $\bar{A}_n$ represents the current inter-slice \gls{RBG} allocation. The action $a_n=[\Delta w_1,\ldots,\Delta w_m]$ applies bounded priority-weight updates $\Delta w_i\in[-\Delta_{\max},\Delta_{\max}]$ to adjust slice precedence in the inter-slice scheduler. The reward quantifies \gls{SLA} adherence in a slice-aware manner, defined as $\mathcal{R}_n=\frac{T_{\mathrm{e}}(n)}{T_{\mathrm{e}}^\star}$ for \gls{eMBB}, $\mathcal{R}_n=\frac{D_{\mathrm{u}}^\star}{D_{\mathrm{u}}(n)}$ for \gls{URLLC}, and $\mathcal{R}_n=\frac{Q^{(5\%)}_{\mathrm{b}}(n)}{\left(Q^{(5\%)}_{\mathrm{b}}\right)^\star}$ for \gls{BE}, where $D_{\mathrm{u}}(n)$ is the measured delay and $D_{\mathrm{u}}^\star$ is the maximum allowable delay, $T_{\mathrm{e}}(n)$ is the instantaneous throughput and $T_{\mathrm{e}}^\star$ is the target throughput, and $Q^{(5\%)}_{\mathrm{b}}(n)$ denotes the $5^{\mathrm{th}}$-percentile throughput with $\left(Q^{(5\%)}_{\mathrm{b}}\right)^\star$ specifying the target value to ensure a minimum service level for cell-edge/low-rate users.

\subsubsection{Supervised Learning-Assisted Self-Healing Using \gls{RAN} Application Activation}

A supervised learning module is designed to recommend the most appropriate network application(s) to initiate based on variations in \glspl{KPI}. The goal is to enable proactive orchestration by detecting KPI degradation patterns and mapping them to predefined application triggers. The module continuously monitors \gls{KPI} measurements and computes variations over a fixed observation window to capture performance trends that may require intervention.

The input vector at time $t$, denoted as $\mathbf{x}_t$, consists of throughput variation $\Delta \mathrm{TP}_t$, packet loss ratio variation $\Delta \mathrm{PLR}_t$, energy efficiency variation $\Delta \mathrm{EE}_t$, channel quality indicator $\mathrm{CQI}_t$, and the aggregate network traffic load $L_t$. Throughput variation $\Delta \mathrm{TP}_t$ is computed as the difference between the current throughput $\mathrm{TP}^{\mathrm{current}}_t$ and the throughput measured $\tau$ time units earlier, $\mathrm{TP}^{\mathrm{past}}_{t-\tau}$. Similar computations are performed for $\Delta \mathrm{PLR}_t$ and $\Delta \mathrm{EE}_t$. The channel quality indicator reflects the overall radio channel condition, while the aggregate network load $L_t$ reflects the current demand on network resources. 

The output label $y_t$ corresponds to the network application or set of applications to be initiated. The output may be represented in a single-label or multi-label format depending on operational requirements. Labels can be generated from expert-defined decision rules or derived from historical operator interventions.

To represent the entire process of the proposed methodology discussed so far, we present Algorithm 4. Note that since the super-agent may trigger the resource-allocation module (Module~3) and the \gls{RAN} application orchestration module (Module~4) sequentially in either order, their independent actions can conflict (e.g., allocating \glspl{RBG} to \gls{eMBB} \glspl{UE} and then activating cell sleeping in the same decision window). To mitigate such cross-module inconsistencies, we adopt a team-learning style information exchange \cite{N37}. When both modules are invoked within the same global decision window, the super-agent forwards the executed action of the first-invoked module as an additional context state to the second module’s internal \gls{MDP}. If Module~3 runs first, its allocation action is appended to Module~4’s state; if Module~4 runs first, its orchestration action (e.g., selected \gls{RAN} application) is appended to Module~3’s state. This conditional action-to-state passing is disabled when only one of the modules is invoked in the window, ensuring coordination only when joint execution is required and preventing unintended coupling across distant time steps.

\begin{algorithm}[t]
\caption{End-to-End Agentic Orchestration via Hierarchical Online Decision Transformer (H-ODT)}
\label{alg:agentic_hodt_e2e}
\small
\begin{algorithmic}[1]
\Require Live KPI stream / network indicators, \gls{SLA} bounds $\mathcal{SLA}$, Predictors $\{F_X\}$, Agent set $\mathcal{A}$, \gls{A-RAG} module $\mathcal{G}$, $\pi_{\theta,\phi}=\{\pi^{\text{control}}_\theta,\pi^{\text{meta}}_\phi\}$, Replay buffer $\mathcal{B}$, Update interval $U$, rollout horizon $H$, context windows $(\omega,K)$

\vspace{2pt}
\State Train an \gls{h-DQN} to generate orchestration demonstrations $\pi_{\text{RL}}$
\State Collect offline trajectories $\mathcal{D}_{\text{RL}}=\{(s_n,g_n,\alpha_{n-\beta},\alpha_n)\}$
\State Initialize \gls{FIFO} replay buffer $\mathcal{B}\leftarrow \mathcal{D}_{\text{RL}}$
\State Pretrain \gls{H-ODT} on $\mathcal{B}$ via supervised sequence modeling
\While{System is running}
    \State Observe network indicators and retrieved analytics $\rightarrow k_n$
    \State Query \gls{A-RAG} for evidence bundle / plan $\rightarrow e_n$
    \State Forecast critical KPIs $f_n=\{\hat{X}(n{+}1)\}$ using $\{F_X\}$

    \If{operator intent arrives}
        \State Parse natural language intent $\rightarrow g=\{m,\delta,z\}$
        \State Set mode $d_n\leftarrow \textsf{HumanIntent}$
    \Else
        \If{\textsf{QoSDrift} detected}
            \State Construct autonomous $g \leftarrow \textsf{AutoGoal}(k_n,f_n,\mathcal{SLA})$
            \State Set mode $d_n\leftarrow \textsf{AutonomousSelfHeal}$
        \Else
            \State \textbf{continue} \Comment{no active intent and no drift}
        \EndIf
    \EndIf

    \If{$d_n=\textsf{HumanIntent}$}
        \State Run \textbf{Slice-aware validator} (Alg. 1)
        \State \textbf{if} \textbf{INVALID} \textbf{then continue}
        \State Run \textbf{\gls{KPI}-centric validator} (Alg. 3)
        \State \textbf{if} \textbf{INVALID} \textbf{then continue}
    \EndIf

    \State \gls{H-ODT} rollout of length $H$ (bi-level stochastic decisions)
    \State Initialize episode buffer $\tau \leftarrow \emptyset$
    \For{$\ell=1$ to $H$}
        \State Form \gls{H-ODT} state $s_\ell=\{z_\ell,k_n,f_n,q_\ell,d_n\}$
        \State Sample $\alpha_{\ell-\beta}\sim \pi^{\text{meta}}_\phi(\cdot\,|\,s_{\ell-\omega:\ell},g)$
        \State Sample $\alpha_\ell\sim \pi^{\text{control}}_\theta(\cdot\,|\,s_{\ell-K:\ell},g,\alpha_{\ell-\beta})$
        \State Execute $\alpha_\ell$; observe $s_{\ell+1}$ and \gls{KPI} measurements
        \State Append $(s_\ell,g,\alpha_{\ell-\beta},\alpha_\ell,s_{\ell+1})$ to $\tau$
    \EndFor

    \State Insert episode $\tau$ into \gls{FIFO} replay buffer $\mathcal{B}$; evict oldest if full
    \State Perform goal-consistent relabeling
    \If{time to update (every $U$ episodes/steps)}
        \State Fine-tune \gls{H-ODT} on $\mathcal{B}$ via supervised sequence modeling
    \EndIf
\EndWhile
\end{algorithmic}
\end{algorithm}

\section{Performance Evaluation}
\label{s5}

This section provides a comprehensive evaluation of the proposed approach. It begins by outlining the computational setup and simulation environment employed in the study. Subsequently, the results are presented to demonstrate the effectiveness of the proposed framework. We first discuss the outcomes of the \gls{LLM} fine-tuning process, which ensures accurate intent and query interpretation. Then, we present the performance of the multi-predictor intent validation technique. Lastly, we highlight the enhancements in \glspl{KPI} achieved through the integration of Agentic \gls{AI} for intent-driven network management.

\subsection{Simulation setup}

The simulation environment in this study features a macro cell surrounded by a dense deployment of small cells operating within a multiple-\gls{RAT} setting. In total, the network serves 60 users. The configuration parameters for the 5G NR and LTE \glspl{RAT} are summarized in Table \ref{sim}.

\begin{table}[!t]
    \centering
    \caption{Simulation and Algorithmic hyperparameter Settings}
    \begin{tabular}{|l|l|}
         \hline
         \textbf{\underline{5G NR}} & \\
         Bandwidth & $50$ and $100$ MHz \\
         Carrier frequency &  $3.5$ and $30$ GHz \\
         Max transmission power & $43$ dBm \\
         Subcarrier spacing & $15$ and $60$ KHz \\
         \hline
         \textbf{\underline{LTE}} & \\
         Bandwidth & $40$ MHz \\
         Carrier frequency & $800$ MHz \\
         Max transmission power & $38$ dBm \\
         Subcarrier spacing & $15$ kHz \\
         \hline
         \textbf{\underline{HRL parameters (Data collection) }} & \\
         Episodic time step, $\tau$ (Numerology = 0) & $20$ TTIs \\
         Episodic time step, $\tau$ (Numerology = 2) & $40$ TTIs \\
         Discount factor, $\gamma$ &  $0.9$ \\
         TTI duration for Numerology 0 and 2 & $1$ ms and $2$ ms\\
         \hline
         \textbf{\underline{DRL parameters}} & \\
         Batch size, Initial exploring steps & $32$, $3000$ \\
         Learning rate ($\alpha$), discount factor ($\gamma$) & $0.5$, $0.9$ \\

         \textbf{\underline{Decision Transformer}} & \\
         Transformer layers & $3$\\
         Attention head & $1$ \\
         Batch size & $64$ \\
         Learning rate & $0.0001$ \\
         
         \hline
    \end{tabular}
    \label{sim}
\end{table}

We consider four types of traffic in this study: video, gaming, voice, URLLC scenario representing vehicle-to-\gls{BS} data transmission, and web browsing traffic. Each traffic type is defined by its packet inter-arrival time, which is 12.5 ms for video, 40 ms for gaming, 20 ms for voice, and 0.5 ms for URLLC, following the specifications in \cite{N44,N45}. For web browsing traffic, there is no single, fixed mean inter-arrival time for data packets. The figure varies depending on the specific activity.  The packet arrival processes differ across traffic types, where video follows a Pareto distribution \cite{N44}, gaming uses a Uniform distribution \cite{N44}, and both voice and URLLC traffic are modeled using Poisson distributions \cite{N45}. For more information on web browsing traffic type, we request the readers to refer to \cite{N44} for more details.  To ensure optimal performance across frequency ranges, we use different antenna configurations and carrier bands. A \gls{ULA} with 64 antennas is employed at the mid-band frequency of $3.5$ GHz to improve coverage and maintain high spectral efficiency due to lower propagation losses. In contrast, a uniform planar array with $128$ antennas is applied at the high band frequency of $30$ GHz to enable advanced beamforming. The bandwidths for these configurations are $60$ MHz for the mid band and $100$ MHz for the high band.

The implementation framework is organized into three integrated layers that jointly enable simulation, learning, and orchestration. The first layer is developed in MATLAB, which simulates the 5G NR and LTE physical and Medium Access Control (MAC) layers, including channel modeling, beamforming, traffic generation, and numerology configuration. This layer provides realistic network data and \glspl{KPI} for higher-level decision-making. The second layer, implemented in Python, hosts all the \gls{AI}-based learning components. It runs the IA\textsuperscript{3}-fine-tuned \gls{LLM} for precise intent and query processing, executes the \gls{H-ODT} for sequential decision-making, and performs multi-model time series forecasting. The third layer is developed using LangGraph, which serves as the Agentic \gls{AI} coordination framework. It orchestrates super-agent workflows across the \gls{RRS}, self-healing, and application orchestration modules, and manages the routing of decisions among different Agentic components.

\subsection{Simulation Results}

\subsubsection{Fine-tuned \gls{LLM} for Intent and Query Processing}

\begin{figure*}[!t]
\centerline{\includegraphics[width=0.8\linewidth]{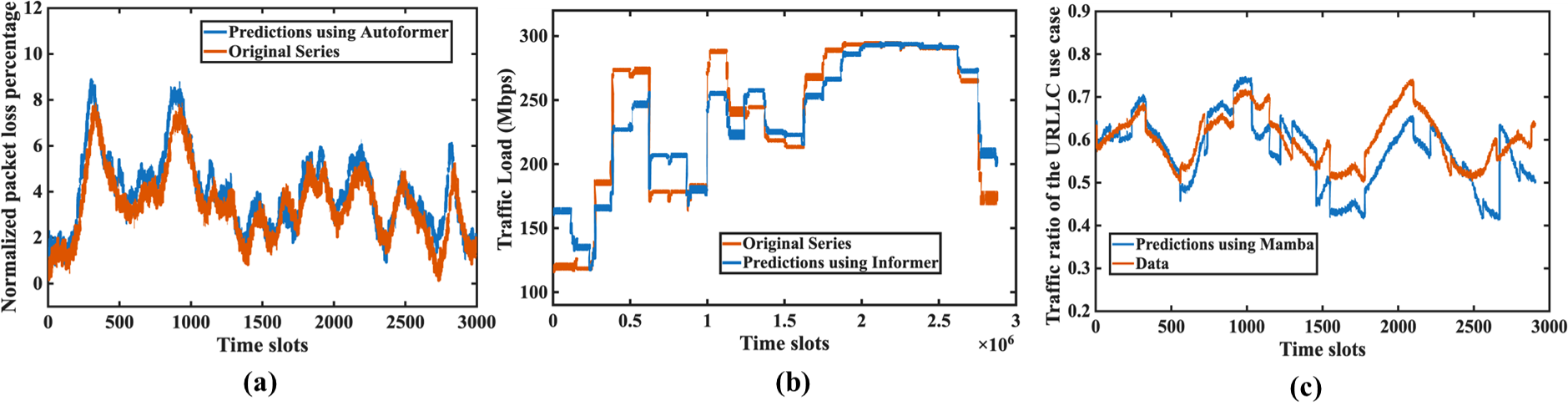}}
\caption{Predicting network parameters using: (a) Autoformer, (b) Informer, and (c) Mamba.}
\label{Fig8}
\end{figure*}

\begin{figure*}[!t]
\centerline{\includegraphics[width=0.9\linewidth]{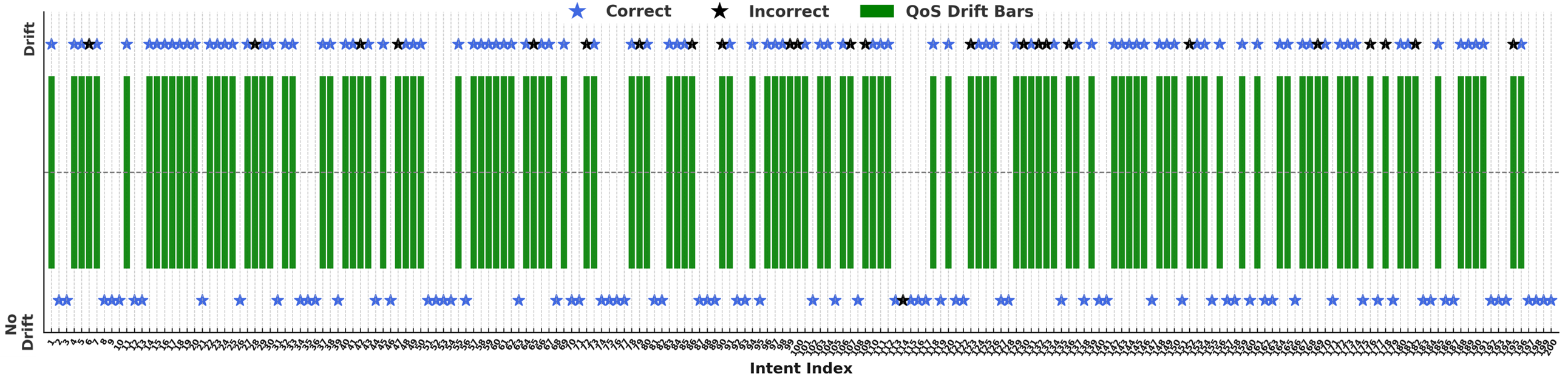}}
\caption{Validation of natural language inputs by operators based on QoS drifts.}
\label{QoS_drift_detect}
\end{figure*}

\begin{figure*}[!t]
\centerline{\includegraphics[width=0.68\linewidth]{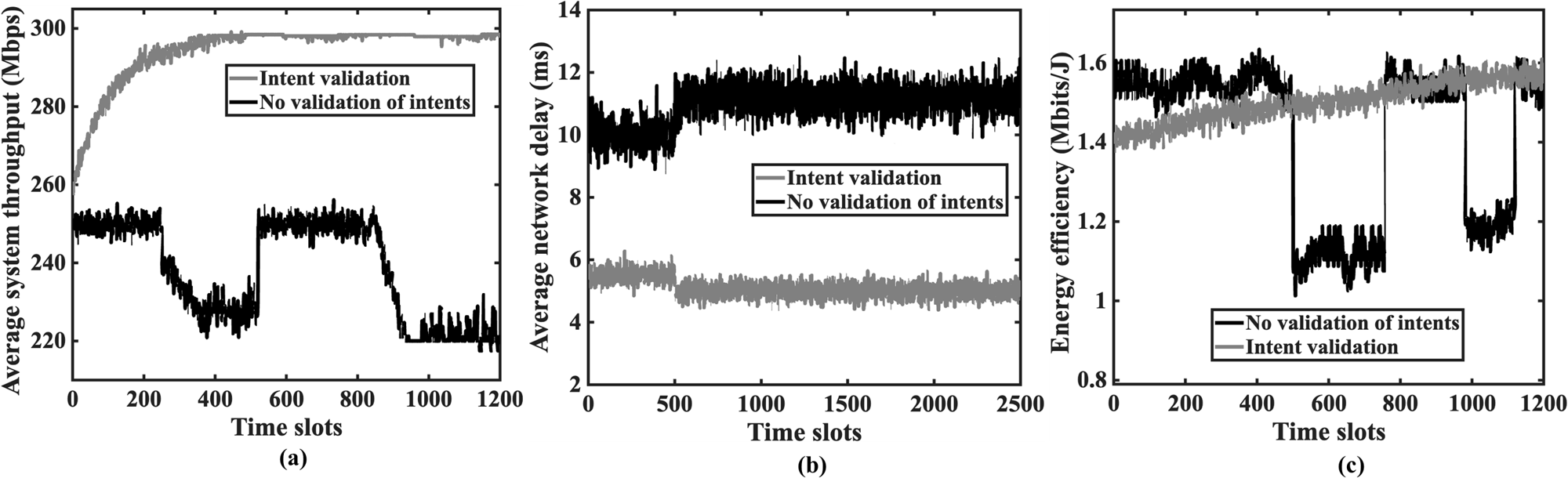}}
\caption{Impact of validating operator commands/intents: (a) Throughput, (b) Network delay and (c) Energy efficiency.}
\label{IV_impact}
\end{figure*}

\begin{table}[!t]
\caption{Performance comparison between IA$^3$-based fine-tuned LLaMA and base LLaMA models.}
\label{tab:llm_eval}
\centering
\small
\begin{tabular}{lcc}
\hline
\textbf{Metric} & \textbf{Base LLaMA} & \textbf{Fine-tuned LLaMA} \\
\hline
BERTScore           & 0.83 & \textbf{0.90} \\
METEOR              & 0.83 & \textbf{0.87} \\
Semantic Similarity & 0.81 & \textbf{0.89} \\
\hline
\end{tabular}
\vspace{-1.0em}
\end{table}

Three crucial metrics are used for performance comparison: \textit{BERTScore} \cite{BERTScore}, \textit{METEOR} \cite{Meteor}, and \textit{semantic similarity} \cite{SS}. These metrics are widely adopted for evaluating generated text quality in tasks such as text generation, summarization, machine translation, and natural language understanding. \textit{BERTScore} measures token-level semantic agreement between generated and reference texts using contextual embeddings and cosine similarity, allowing evaluation beyond exact word overlap. \textit{METEOR} captures lexical quality by considering precision, recall, synonym matching, and word order. \textit{Semantic similarity} measures the degree to which two texts convey the same meaning at the sentence level, even when their wording differs.

As summarized in Table~\ref{tab:llm_eval}, the IA$^3$-based fine-tuned LLaMA model consistently outperforms the base model across all metrics by improving BERTScore from 0.83 to 0.90, METEOR from 0.83 to 0.87, and semantic similarity from 0.81 to 0.89. These gains indicate enhanced semantic fidelity, lexical accuracy, and intent preservation, which directly translate into more reliable intent interpretation and reduced risk of erroneous downstream orchestration decisions in the proposed Agentic framework.

\subsubsection{Intent Validation} 

The core of our intent validation is successfully predicting key network parameters such as the traffic-mix ratio of a traffic class, the aggregate traffic load, power consumption, and the packet loss percentage. We maintain an inventory of multiple time-series predictors, and the super agent selects the most suitable model based on the underlying characteristics of the data. For example, the traffic-mix ratio is typically a highly non-stationary and bursty sequence with sharp regime changes caused for \gls{URLLC} arrivals. To capture these short-range spikes and transient dynamics with low-latency inference, the super agent prioritizes a Mamba-based predictor for this category. In Fig. \ref{Fig8}c, we present the prediction output using Mamba. Particularly, we have used Mamba4Cast \cite{N50} to perform prediction. 

In contrast, traffic-load prediction operates over longer horizons, often at the scale of an entire day and exhibits strong diurnal periodicity with smooth ramp-up and plateau regions. For such data, the Informer predictor is preferred, as its encoder–decoder sparse attention can effectively align multi-day historical patterns with future trajectories when calendar-based exogenous features are present.  In Fig. \ref{Fig8}b, we present the prediction output using Informer \cite{N40}. The super agent, therefore, routes each validation request to the model that best matches the statistical properties of the target parameter to ensure stable forecasts across both short-term bursty signals and long-term seasonal trends. These forecasts are then evaluated against operator goals and domain constraints to determine whether the intent is feasible, partially feasible, or requires additional corrective actions before execution.

The fine-tuned super agent employs multiple predictors to forecast different network parameters for intent validation using either the slice-aware validation algorithm (Algorithm~1) or the KPI-centric validation algorithm (Algorithm~3). More than 500 intents were evaluated prior to deployment, achieving an average validation accuracy of 88.5\%. Fig.~\ref{QoS_drift_detect} illustrates the drift-detection performance over 200 representative intents, where 185 intents were correctly classified. Green bars indicate intents that triggered QoS drift, while blue stars denote correct decisions and black stars denote missed detections. The high density of correct detections and the low number of missed events demonstrate the effectiveness of the proposed Agentic AI framework in adaptive model selection, contextual reasoning, and real-time validation.

Without intent validation, misaligned intents impact KPIs. Fig. \ref{IV_impact} illustrates fluctuating throughput (\ref{IV_impact}a), delay (\ref{IV_impact}b), and energy efficiency (\ref{IV_impact}c) without intent validation. With intent validation, KPI curves are smoother, ensuring stable performance.

\subsection{Intent Execution}

\begin{figure}[!t]
\centerline{\includegraphics[width=0.9\linewidth]{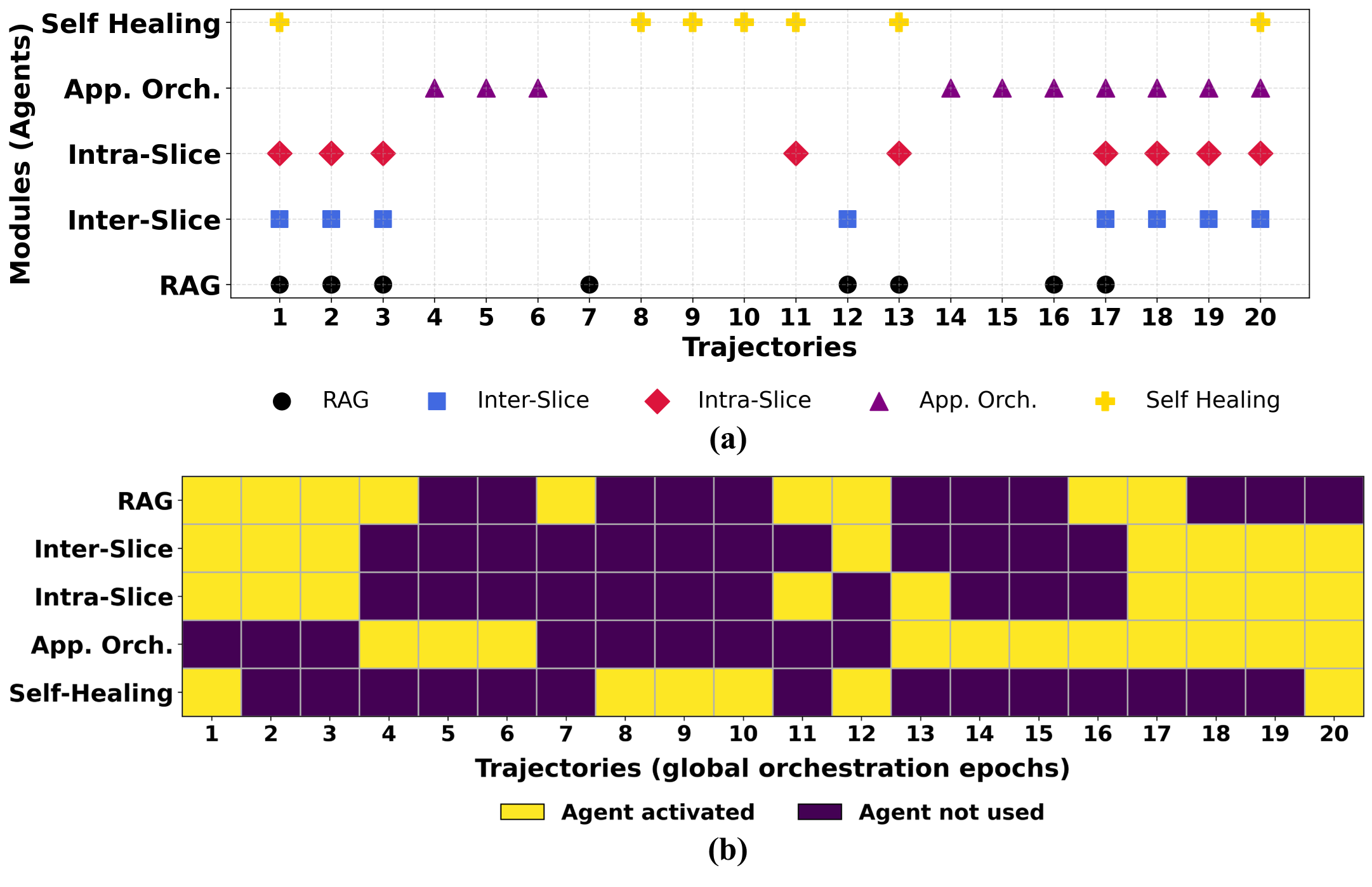}}
\caption{Selective event-driven agent activation across orchestration epochs in the proposed control framework: (a) scatter plot showing which functional agent is invoked at each epoch, and (b) binary heatmap indicating whether each agent is activated or not activated over time.}
\label{agentic_orch}
\vspace{-1.2em}
\end{figure}

\begin{figure*}[!t]
\centerline{\includegraphics[width=0.88\linewidth]{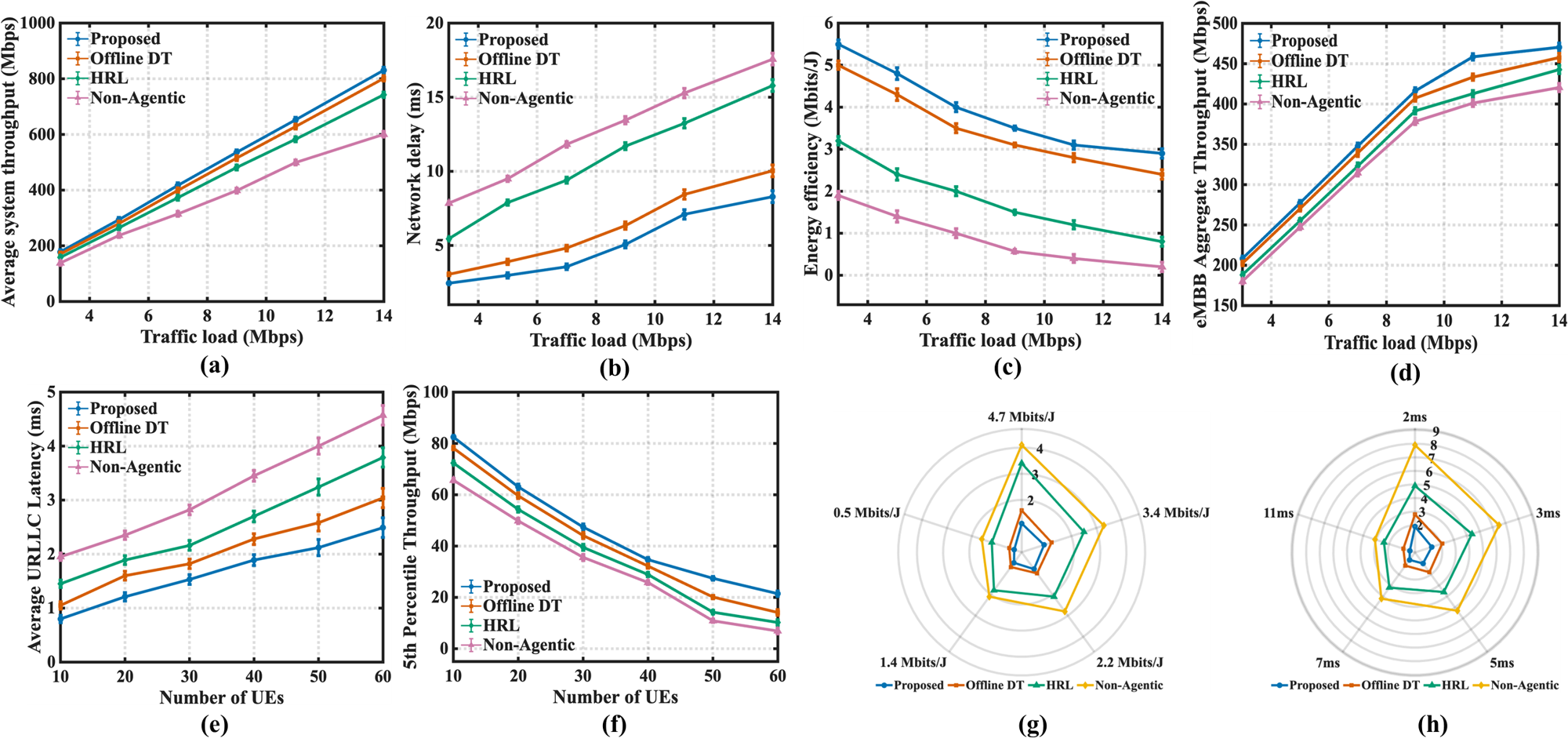}}
\caption{Performance analysis in terms of:
(a) Average network delay, (b) Average system throughput, and (c) energy efficiency, (d) eMBB Aggregate throughput, (e) URLLC latency, (f) 5\textsuperscript{th} percentile throughput. Spider plots illustrating deviation from operator-defined goals:  (g) energy-efficiency and (h) latency; smaller polygons indicate closer goal satisfaction.}
\label{global_kpi_figs}
\vspace{-1.2em}
\end{figure*}

Our proposed system is based on Agentic \gls{AI} where an \gls{H-ODT} works as a backbone to make decisions regarding which agents to sequentially initiate. Fig. \ref{agentic_orch}a presents a scatter-based visualization of the event-driven orchestration behavior of the proposed Agentic AI framework. Each point along the horizontal axis corresponds to a global orchestration trajectory at which the super-agent evaluates intent feasibility, predicted KPIs, and validation outcomes, while the vertical axis enumerates the available functional agents. A marker indicates that a particular agent is activated at that trajectory. The sparse distribution of markers demonstrates that the super-agent does not continuously invoke all modules; instead, it selectively triggers reasoning, scheduling, orchestration, or self-healing actions only when required. Periods with RAG-only activation indicate stable network conditions where no control intervention is necessary.
Fig. \ref{agentic_orch}b provides a binary heatmap representation of the Agentic orchestration process. This visualization reinforces that the proposed Agentic AI framework minimizes unnecessary control actions by activating agents only when warranted by intent validation and predicted \gls{KPI} deviations, thereby reducing orchestration overhead while maintaining stable and intent-compliant RAN operation.

So far, we have presented results associated with the operator providing input in natural language to guide the optimization processes using Agentic AI. Also, how the natural language input can be validated, and the impact on the crucial KPIs. However, the main goal of this work is to optimize \gls{RAN}. In order to do that, we provide performance graphs associated with some crucial \glspl{KPI} that include system throughput, network latency, and energy efficiency. To show the superiority of our proposed Agentic AI-based methodology, we compare the performance achievement with three baselines. 

\textbf{1) Baseline 1:} We include an Offline \gls{DT} as a baseline. This baseline preserves the same Agentic AI architecture, agent interactions, state and action representations, and execution modules as the proposed framework. The only distinction is that the super agent is controlled by a purely offline \gls{DT} which is trained on historical orchestration trajectories, with no access to online observations, trajectory updates, or corrective feedback during deployment. 

\textbf{1) Baseline 2:} We also consider an \gls{HRL} baseline to represent a classical alternative for long-horizon decision-making in intent-driven \gls{RAN} management. This baseline retains the same Agentic \gls{AI} architecture, agent set, state and action spaces, and execution modules as the proposed framework.

\textbf{1) Baseline 3:} We include a non-Agentic module selection baseline that represents a classical heuristic approach to intent execution without sequential planning or agent coordination. For a given intent, this baseline selects a subset of execution modules once by iteratively adding the module that yields the largest marginal estimated improvement per unit cost, subject to a fixed budget constraint.

Fig.~\ref{global_kpi_figs}a, Fig.~\ref{global_kpi_figs}b, and Fig.~\ref{global_kpi_figs}c present the system-wide network performance comparison under increasing traffic load. Fig.~\ref{global_kpi_figs}a illustrates the average system throughput performance. Relative to the Offline \gls{DT}, \gls{HRL}, and non-Agentic baselines, the proposed \gls{H-ODT}-based Agentic intelligence improves the average system throughput by approximately $4.3\%$, $11.8\%$, and $31.5\%$, respectively, across all traffic load scenarios. Furthermore, as shown in Fig.~\ref{global_kpi_figs}b, the proposed method delivers substantial gains in energy efficiency, achieving approximately $13.4\%$ improvement over offline \gls{DT}, about $1.4$ times improvement over HRL, and more than $5$ times improvement over non-Agentic control across the evaluated traffic loads. Finally, as illustrated in Fig.~\ref{global_kpi_figs}c, compared with the Offline \gls{DT}, \gls{HRL}, and non-Agentic baselines, the proposed \gls{H-ODT}-based Agentic intelligence achieves approximately $20.5\%$, $55.0\%$, and $62.7\%$ reduction in average network latency, respectively, across all traffic load conditions.

Fig. \ref{global_kpi_figs}d, \ref{global_kpi_figs}e and \ref{global_kpi_figs}f illustrate the slice-level \gls{QoS} performance of the network under increasing user density and traffic load. Fig. \ref{global_kpi_figs}d shows that the proposed approach also improves \gls{eMBB} aggregate throughput by up to $11.9$\%, demonstrating that enhanced tail reliability and \gls{URLLC} protection do not come at the expense of overall system capacity. As shown in Fig. \ref{global_kpi_figs}e, the proposed Agentic \gls{AI} framework consistently achieves lower average \gls{URLLC} latency as the number of \glspl{UE} increases, with up to 45\% latency reduction at high load compared to the non-Agentic baseline. Finally, Fig. \ref{global_kpi_figs}f presents the 5th-percentile throughput, reflecting tail-user performance. The proposed method achieves approximately $2$-$3$ times higher tail throughput than the baseline schemes under heavy load, indicating significantly improved fairness and reduced user starvation. Fig. \ref{global_kpi_figs}g and Fig. \ref{global_kpi_figs}h present spider plots illustrating the deviation from operator-defined performance goals. The radius denotes deviation from the target goal (lower is better). Fig. \ref{global_kpi_figs}g shows energy-efficiency targets, and Fig. \ref{global_kpi_figs}h presents latency targets. Across all operating points, the proposed Agentic \gls{AI} framework consistently exhibits the smallest polygon area, indicating the lowest deviation from the desired goals compared to the baselines. Similar performance can be observed for the throughput goals as well. 

The observed performance improvements stem from three factors: Agentic orchestration, online hierarchical decision intelligence, and wireless-aware control design. From the Agentic perspective, we do not rely on a single monolithic controller; instead, the super-agent coordinates specialized functional agents (e.g., retrieval/reasoning, inter-slice scheduling, intra-slice scheduling, application orchestration, and self-healing) and activates them only when needed. This reduces unnecessary signaling, computation, and reconfiguration overhead and thereby improves energy efficiency and latency stability. From the H-ODT perspective, our super-agent is not frozen like an offline decision transformer; it continuously incorporates newly collected trajectories, which mitigates distribution shift under changing traffic loads and improves long-horizon planning by learning effective sequences of actions rather than short-term KPI reactions. From the wireless/RAN perspective, our hierarchical separation between inter-slice and intra-slice control better matches practical RAN dynamics. It strengthens slice isolation, protects \gls{URLLC} latency while preserving \gls{eMBB} capacity, and improves tail-user fairness (e.g., 5th-percentile throughput) by reducing starvation under heavy load. Moreover, predictive intent validation blocks risk-prone actions near congestion regimes, preventing queue build-up and \gls{KPI} collapse. As a result, we achieve higher throughput, lower delay, higher energy efficiency, improved worst-user performance, and consistently smaller deviation from operator-defined goals, as reflected by the reduced polygon areas in the spider plots.

\begin{figure}[!t]
\centerline{\includegraphics[width=0.9\linewidth]{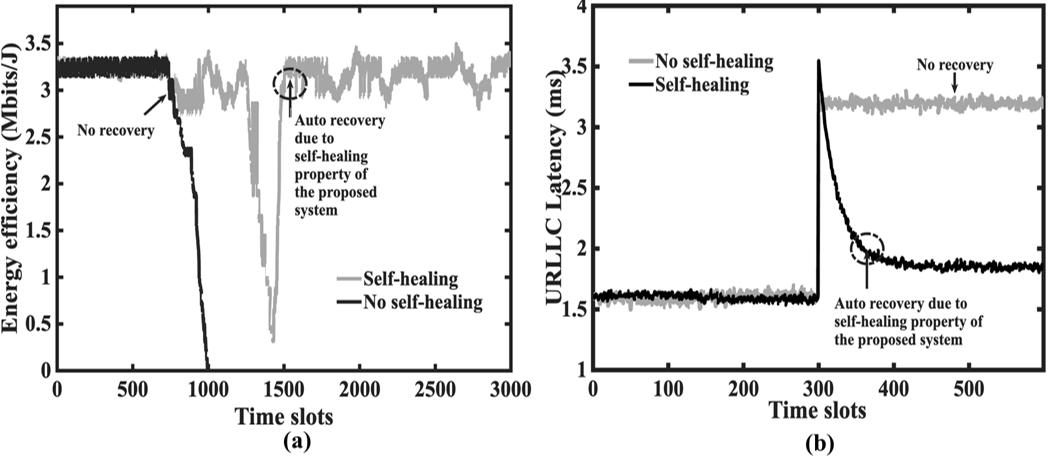}}
\caption{Performance analysis in terms of:
(a) effect of self-healing property on energy efficiency, and (b) effect of self-healing property on URLLC latency.}
\label{self_heal_kpi_figs}
\vspace{-1.2em}
\end{figure}

Fig. \ref{self_heal_kpi_figs}a and Fig. \ref{self_heal_kpi_figs}b illustrate the self-healing capability of the proposed Agentic \gls{AI} framework under performance-degrading events. In Fig. \ref{self_heal_kpi_figs}a, the energy efficiency remains stable around its nominal operating region until a sudden degradation occurs, causing a sharp drop due to an injected fault or adverse network condition. This drop is highlighted as a performance-degrading event. Following the detection of this degradation, the self-healing mechanism is autonomously triggered, resulting in a rapid recovery of energy efficiency back to its pre-fault level. A similar behavior is observed in Fig. \ref{self_heal_kpi_figs}b for throughput performance. The throughput experiences a pronounced decline during the degradation event, reflecting the immediate impact of the fault on network operation. After self-healing actions such as corrective resource reallocation and application orchestration are executed, the throughput gradually recovers and stabilizes near its original operating point. 

To quantitatively evaluate the robustness and autonomous self-healing capability of the proposed Agentic framework, we manually injected $127$ heterogeneous performance-degrading events, including abrupt traffic surges, aggressive cell sleeping, and resource perturbations. These events emulate realistic operational disturbances encountered in practical \gls{RAN} deployments.

\begin{figure}[!t]
\centerline{\includegraphics[width=0.91\linewidth]{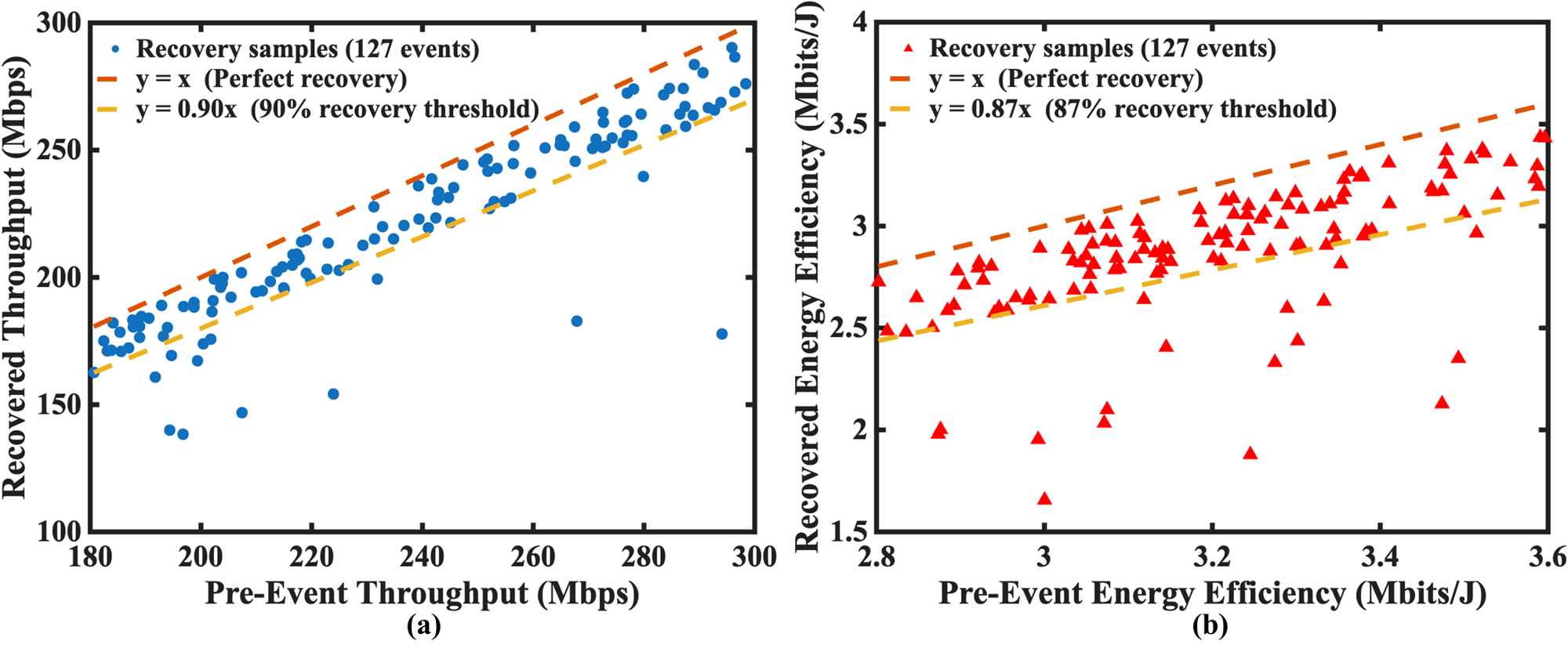}}
\caption{Self-healing recovery performance under injected degradation events: (a) throughput, and (b) energy efficiency.}
\label{sh_tp_ee}
\vspace{-1.2em}
\end{figure}

For each degradation event $i$, the recovery ratio is defined as: $\rho_i = \frac{\mathrm{KPI}_i^{\mathrm{rec}}}{\mathrm{KPI}_i^{\mathrm{pre}}}$, where $\mathrm{KPI}_i^{\mathrm{pre}}$ denotes the steady-state operating point prior to the disturbance, and $\mathrm{KPI}_i^{\mathrm{rec}}$ represents the stabilized performance after the self-healing mechanism converges. Fig. \ref{sh_tp_ee}a and \ref{sh_tp_ee}b illustrate the recovery behavior for throughput and energy efficiency. The same behavior can be observed for network delay as well. Each marker corresponds to a single injected degradation event, with the horizontal and vertical axes indicating the pre-event and recovered KPI values. The upper dashed reference line ($y=x$) represents perfect recovery, while the lower dashed line indicates the minimum acceptable recovery threshold $y=\chi x$, where $\chi$ is selected based on the service-level objective of each \gls{KPI}. For throughput, more than $90\%$ of the events exceed the recovery threshold ($\chi=0.90$), demonstrating that the orchestration policy reliably restores near-nominal capacity even after severe performance collapse. The tight clustering of points near the perfect-recovery line further indicates low residual degradation and stable convergence. For energy efficiency, approximately $87\%$ of the events satisfy the recovery criterion ($\chi=0.87$), confirming that the framework effectively rebalances power allocation and cell activation decisions following disruptive network dynamics while avoiding excessive operational overhead. Similar effect is observed for the network delay ($93\%$ of the events satisfy the recovery criterion)

\section{Conclusions}
\label{s6}  

This paper proposed an Agentic \gls{AI} framework for intent-driven RAN automation in \gls{6G}, where a super agent powered by an Online Decision Transformer orchestrates multiple specialized agents for resource allocation, network application execution, and self-healing. By integrating Agentic Retrieval-Augmented Generation and a bi-level intent validation mechanism, the framework enables safe, context-aware, and autonomous decision-making even in the absence of human operators. Simulation results show that the proposed approach outperforms both non-Agentic and learning-based baselines, achieving up to $32.9$\% higher throughput, up to $60.9$\% lower delay, and up to threefold improvement in energy efficiency, while rejecting performance-degrading intents with $88.5$\% accuracy and restoring approximately $90$\% of pre-degradation performance through self-healing. These results highlight Agentic AI with online decision intelligence as a practical and scalable pathway toward zero-touch RAN management in future \gls{6G} networks.

\section*{Acknowledgment}

This work has been supported by MITACS, Ericsson Canada, and the NSERC Canada Research Chairs program.

\section*{Appendix A}

The dataset for the LLM fine-tuning was grounded in realistic 5G/6G orchestration scenarios involving intent recognition, reasoning, and agent coordination. A subset of intents and reasoning traces were empirically tested within the simulated Agentic AI environment to confirm that the model’s predicted responses induced meaningful network actions. Some examples of this dataset are presented in Table \ref{table_dataset}.

\begin{table}[H]
\centering
\scriptsize
\caption{Example query--response pairs for IA$^3$-based LLM fine-tuning}
\label{table_dataset}
\setlength{\tabcolsep}{2pt}
\renewcommand{\arraystretch}{1.05}
\begin{tabularx}{\columnwidth}{|>{\raggedright\arraybackslash}X|>{\raggedright\arraybackslash}X|}
\hline
\textbf{Query} & \textbf{Response} \\ \hline

\multicolumn{2}{|c|}{\textbf{Intent--Action Reasoning}} \\ \hline
Increase URLLC reliability by 20\%. 
& Intent detected: URLLC reliability improvement. Retrieve slice metrics, identify latency-sensitive users, and activate the intra-slice scheduler to prioritize low-latency UEs. Expected delay $\leq 1$ ms. \\ \hline

Boost overall throughput by 10\%. 
& Intent type: throughput increase. The inter-slice agent allocates more RBGs to the eMBB slice and triggers beamforming to improve SINR in high-load sectors. \\ \hline

Reduce packet drop rate in the eMBB slice. 
& Intent type: reliability improvement. Use RAN analytics to detect buffer overflow, activate traffic steering to a neighboring 5G cell, and update the orchestration log. \\ \hline

\multicolumn{2}{|c|}{\textbf{Multi-Database Retrieval (A-RAG)}} \\ \hline
What are the O-RAN specifications for traffic steering? 
& Retrieved from O-RAN WG1: traffic steering adjusts UE routing through E2 control while remaining compliant with A1 policy constraints. \\ \hline

What historical intents improved energy efficiency last week? 
& Three prior intents involving power allocation and cell sleeping were found. Average gain: 17\%. Best policy: joint power control with eMBB offloading. \\ \hline

\multicolumn{2}{|c|}{\textbf{Self-Healing and QoS Drift}} \\ \hline
Throughput dropped by 12\%, no human intent detected. 
& Autonomous trigger: activate beamforming xApp to restore SINR. Monitor KPI updates and revert if throughput remains below baseline after 3 TTIs. \\ \hline

Delay deviation detected in URLLC slice. 
& Alert: latency is 2.3$\times$ above target. Activate inter-slice reallocation and assign extra RBGs from the BE slice to URLLC. \\ \hline

All cells are stable except Cell~3, showing a drop in energy efficiency. 
& Self-healing response: reduce transmission power by 10\% and enable cell sleeping for Cell~3. Expected energy-efficiency gain: 14\%. \\ \hline

\multicolumn{2}{|c|}{\textbf{Agentic Coordination}} \\ \hline
Current throughput is 290~Mbps; increase by 5\%. 
& The super agent initiates the inter-slice agent for RBG reallocation, while the orchestration agent triggers beamforming. The new throughput is then validated through RAG. \\ \hline

Performance degradation in Cell~2, BE slice. 
& The super agent identifies the drift source, triggers intra-slice reallocation, and updates the intent-history database through RAG. \\ \hline

Validate the last orchestration effect on delay reduction. 
& The super agent queries RAG for KPI updates. Retrieved latency improved by 32\%, so the intent is marked as fulfilled. \\ \hline

\end{tabularx}
\end{table}

\bibliographystyle{IEEEtran}
\bibliography{reference.bib}

\end{document}